\documentclass[aip,pop, reprint]{revtex4-1}
\usepackage{bm}
\usepackage{amsmath,amsfonts,amssymb}
\usepackage{blkarray}
\usepackage{color}

\newcommand{\vq}{\mathbf}
\newcommand{\vc}{\mathbf}

\draft 

\begin{document}


\title{Foundations of magnetohydrodynamics} 


\author{Jarett LeVan}
\email[]{jarettl@umich.edu}
\author{Scott D. Baalrud}
\email[]{baalrud@umich.edu}
\affiliation{Nuclear Engineering \& Radiological Sciences, University of Michigan, Ann Arbor, Michigan 48109, USA}

\date{\today}

\begin{abstract}
In this tutorial, a derivation of magnetohydrodynamics (MHD) valid beyond the usual ideal gas approximation is presented. Non-equilibrium thermodynamics is used to obtain conservation equations and linear constitutive relations. 
When coupled with Maxwell's equations, this provides closed fluid equations in terms of  material properties of the plasma, described by the equation of state and transport coefficients. 
These properties are connected to microscopic dynamics using the Irving-Kirkwood procedure and Green-Kubo relations. Symmetry arguments and the Onsager-Casimir relations allow one to vastly simplify the number of independent coefficients. Importantly, expressions for current density, heat flux, and stress (conventionally Ohm's law, Fourier's law, and Newton's law) take different forms in systems with a non-ideal equation of state. 
The traditional form of the MHD equations, which is usually obtained from a Chapman-Enskog solution of the Boltzmann equation, corresponds to the ideal gas limit of the general equations. 
\end{abstract}


\maketitle 

\section{Introduction}
\label{sec-intro}
Magnetohydrodynamic (MHD) equations are a cornerstone of plasma physics, providing a bridge between electromagnetism and fluid dynamics. For nearly a century, they have been employed to study complex behavior in plasmas, providing insight into fusion experiments~\cite{Freidberg_1982}, stellar atmospheres~\cite{Alfven_1942}, and magnetic reconnection~\cite{recon_rev}, among a wide range of other phenomena~\cite{beresnyak_2019, Hawley_2003, burlaga1984mhd}. They are often obtained by applying the Chapman-Enskog procedure to the Boltzmann equation~\cite{chapman_cowling,braginskii_review}, which is valid only for dilute plasmas that satisfy an ideal gas equation of state.  

In this tutorial, a more general derivation of MHD equations is presented. This is achieved by applying non-equilibrium thermodynamics,~\cite{degroot_mazur} which provides a macroscopic framework for describing irreversible processes, to plasmas. 
First, conservation laws for mass, momentum and energy are derived from macroscopic arguments. Non-equilibrium thermodynamics is then used to develop expressions for the heat flux, current density, and viscous pressure tensor. In particular, the entropy production rate is analyzed and used to establish linear constitutive relations between the thermodynamic forces and fluxes. 
Symmetry arguments and the Onsager-Casimir relations are also discussed,~\cite{Onsager_1931,Casimir_1945} which allow one to show that of the $225$ coefficients that appear in the most general form of the linear constitutive relations, only $16$ are independent. 
This provides the structure of the MHD equations, but does not offer a way to compute the 16 transport coefficients or the equation of state.

Computing these values requires a connection with microscopic dynamics. 
Here, it is shown how the Irving-Kirkwood procedure~\cite{Irving_Kirkwood}, in combination with the Green-Kubo relations~\cite{Green_1954, kubo_1957}, enable one to evaluate them from particle trajectories in an equilibrium system. 
This provides a closed and self-consistent description of MHD so long as there is a means to model the particle trajectories. 
There may be many ways to do this, including experiments and approximate models, but a popular and accurate means is molecular dynamics (MD) simulations~\cite{transport_workshop}. 

In comparison with the standard method to derive MHD based on a Chapman-Enskog solution of the Boltzmann equation, this method has the advantage of generality. 
It applies to a wide range of Coulomb coupling and electron degeneracy regimes, making only the single fluid and local thermodynamic equilibrium assumptions that define the MHD limit. 
It captures transport processes that are not a part of traditional MHD, including bulk viscosity and thermodynamic gradients that arise when the plasma is not ideal. 
However, it does not provide simple formulas for the transport coefficients, as the traditional solution does. 
Instead these must be experimentally measured, or computed from some other means of modeling the microscopic dynamics, such as MD. 
Of course, the general and traditional approaches are consistent in the weakly coupled limit (i.e., the dilute ideal-gas limit) so the usual Chapman-Enskog solutions may be used to the model the transport coefficients in this regime. 
Differences arise when the plasma is sufficiently dense or cold to be non-ideal. 

In fact, this tutorial is motivated, in part, by an apparent inconsistency in a common approach to modeling transport in dense plasmas. It is common to evaluate non-ideal equations of state and transport coefficients, then insert the values into the traditional structure of the MHD equations. 
Here, we stress that dense plasma effects not only change the equation of state and transport coefficients, but also the structure of the MHD equations. 
For instance, in a non-ideal plasma Ohm's law and Fourier's law depend on gradients of the electron chemical potential, as opposed to the commonly used pressure gradient. 
Another instance is the presence of bulk and cross viscosity terms in Newton's law, which arise particularly if a significant fraction of molecular species are present,~\cite{LeVan_2024,LeVan_2025} but vanish in the monatomic ideal gas limit. 

The remainder of this tutorial is organized as follows. 
Section~\ref{sec-macro} describes macroscopic considerations, deriving the structure of the MHD equations from conservation laws, development of linear constitutive relations from considerations of entropy production, and the symmetry properties of these equations. 
Section~\ref{sec:micro} describes the connection between transport coefficients and equation of state properties that arise in the macroscopic equations with the microscopic dynamics of particle trajectories. 
This utilizes the Irving-Kirkwood procedure and Green-Kubo relations. 
Finally, Sec.~\ref{sec:compare} provides a discussion and comparison of the general equations with those obtained from other methods, including the standard MHD equations in the ideal gas limit.

\section{Macroscopic considerations~\label{sec-macro}}
Hydrodynamic equations describe systems over length and time scales much longer than those that govern the underlying microscopic dynamics. The domain of a system is divided into small volumes called fluid elements, and it is the goal of hydrodynamics to determine the density, flow velocity, and temperature associated with each element. For the theory to be valid, the length of each fluid element in any given direction must be larger than the mean free path of its constituent particles, and the fluid variables must change significantly only on timescales longer than the mean collision time. Moreover, though the system as a whole need not be in equilibrium, it is assumed that each fluid element satisfies the fundamental equations of thermodynamics. This is known as the assumption of local thermodynamic equilibrium.~\cite{degroot_mazur} To begin constructing fluid equations, one considers how the fluid variables must change in order to satisfy conservation of mass, momentum, and energy. 

\subsection{Conservation equations}
Consider a plasma property $A$ within an arbitrary volume $\mathcal{V}$. If the property is conserved, its density $a=A/\mathcal{V}$ 
can change only if there is a net flux of $A$ across the volume boundary $\Omega$. It follows that
\begin{equation}
    \frac{d}{dt} \int_{\mathcal{V}} a\, d\mathcal{V} = \int_{\mathcal{V}} \frac{\partial a}{\partial t} d\mathcal{V} = - \int_\Omega \bm{\mathcal{J}}_A \cdot d\bm{\Omega}
\end{equation}
where $\bm{\mathcal{J}}_A$ is the flux of $A$. Application of Gauss's theorem yields
\begin{equation}
    \int_\mathcal{V} \left[ \frac{\partial a}{\partial t} + \nabla \cdot \bm{\mathcal{J}}_A \right] d\mathcal{V} = 0.
\end{equation}
Since this expression must hold for any volume $\mathcal{V}$, the integral can be dropped altogether, leaving
\begin{equation}
\label{eq:general-conservation}
    \frac{\partial a}{\partial t} = - \nabla \cdot \bm{\mathcal{J}}_A.
\end{equation}
This is the general form of a conservation equation, and it can be used to construct expressions for the conservation of mass, momentum, and energy in a plasma. 

\subsubsection{Conservation of mass}
In the absence of reactions that convert species from one type to another, the mass $m_\alpha$ of any species $\alpha$ must be conserved. Application of Eq.~(\ref{eq:general-conservation}) yields
\begin{equation}
\label{eq:mass-alpha}
    \frac{\partial \rho_\alpha}{\partial t} = - \nabla \cdot (\rho_\alpha \vq{V}_\alpha)
\end{equation}
where $\rho_\alpha = m_\alpha/\mathcal{V}$ is the mass density of species $\alpha$, $\vq{V}_\alpha$ is the net velocity of $\alpha$, and $\bm{\mathcal{J}}_{M_\alpha} = \rho_\alpha \vq{V}_\alpha$ is the total mass flux of $\alpha$. This expression may be summed over all $N_S$ species to obtain
\begin{equation}
    \frac{\partial \rho}{\partial t} = - \nabla \cdot (\rho \vq{V})
\end{equation}
where $\rho = \sum_\alpha \rho_\alpha$ is the total mass density and $\vq{V} = \sum_\alpha \rho_\alpha \vq{V}_\alpha/\rho$ is the fluid's center of mass velocity. Introducing the Lagrangian derivative $d/dt = \partial/\partial t + \vq{V} \cdot \nabla$, the total mass conservation equation may alternatively be written 
\begin{equation}
\label{eq:final-mass}
    \frac{1}{\rho} \frac{d\rho}{dt} = - \nabla \cdot \vq{V}.
\end{equation}
This is the form of the mass conservation equation that will be used below. However, before advancing to momentum conservation, it is convenient to define a diffusive mass flux
\begin{equation}
\label{eq:diff-flux}
    \vq{d}_\alpha = \rho_\alpha (\vq{V}_\alpha - \vq{V}).
\end{equation}
With this definition and the Lagrangian derivative, the mass conservation of species $\alpha$ from Eq.~(\ref{eq:mass-alpha}) can be written 
\begin{equation}
\label{eq:partial-mass}
    \rho \frac{dc_\alpha}{dt} = - \nabla \cdot \vq{d}_\alpha.
\end{equation}
where $c_\alpha = \rho_\alpha / \rho$ is the mass fraction of species $\alpha$. This form will prove to be useful when studying entropy production.

\subsubsection{Conservation of momentum}
Momentum of matter and field must be conserved, so
\begin{equation}
\label{eq:mom-start}
    \frac{\partial}{\partial t} \left(\rho \vq{V} + \epsilon_0 \vq{E} \times \vq{B} \right) = - \nabla \cdot \left( \rho \vq{V} \vq{V} + \vq{P} - \bm{\mathcal{T}}_\textrm{M} \right)
\end{equation}
where $\rho \vc{V}$ is the plasma momentum, $\vq{E}$ is the electric field, $\vq{B}$ is the magnetic field, and $\epsilon_o \vc{E}\times \vc{B}$ is the Poynting flux describing the electromagnetic field momentum. The corresponding fluxes are the convection of plasma momentum $\rho \vc{V} \vc{V}$, the plasma pressure tensor $\vq{P}$, and for the field, the Maxwell stress tensor  
\begin{equation}
\bm{\mathcal{T}}_\textrm{M} = \epsilon_o \biggl( \vq{E} \vq{E} - \frac{1}{2} E^2 \mathcal{I} \biggr) + \frac{1}{\mu_o} \biggl( \vc{B} \vq{B} - \frac{1}{2} B^2 \mathcal{I} \biggr).
\end{equation}
The divergence of Maxwell's stress tensor may be evaluated directly from Maxwell's equations as
\begin{equation}
\label{eq:maxwell-stress}
    \nabla \cdot \bm{\mathcal{T}}_\textrm{M} = \rho z \vq{E} + \vq{J} \times \vq{B} + \epsilon_0 \frac{\partial}{\partial t}\vq{E} \times \vq{B}
\end{equation}
where $\vq{J}$ is the total current density, and $\rho z$ is the charge density.

To relate the current density to species in the system, define 
\begin{equation}
    z_\alpha = \frac{q_\alpha}{m_\alpha} \quad \mathrm{and} \quad   z = \frac{1}{\rho}\sum_\alpha^{N_s} \rho_\alpha z_\alpha
\end{equation}
where $q_\alpha$ is the net charge of particles of species $\alpha$, so that $z_\alpha$ is the charge per unit mass of species $\alpha$ and $z$ is the total charge per unit mass. 
Note that in a partially ionized plasma, the charge of any particular species $\alpha$ may be zero. 
Charge-neutral species contribute to the mass density, but not the charge density. 
The standard MHD assumption of quasi-neutrality will be made, which sets $z = 0$.
The total current density is then defined as
\begin{equation}
    \vq{J} = \sum_\alpha^{N_s} z_\alpha \rho_\alpha \vq{V}_\alpha = z \rho \vq{V} + \vq{j}
\end{equation}
where $z \rho \vq{V}$ is the convective current and
\begin{equation}
    \vq{j} = \sum_\alpha z_\alpha \vq{d}_\alpha
\end{equation}
is the diffusive current. 
Since $z = 0$, the total current density is equal to the diffusive current: $\vq{J} = \vq{j}$.
Insertion of Eq.~(\ref{eq:mom-start}) into Eq.~(\ref{eq:maxwell-stress}) yields the familiar momentum conservation equation
\begin{equation}
\label{eq:final-mom}
    \rho \frac{d\vq{V}}{dt} = - \nabla \cdot \vq{P} +  \vq{j} \times \vq{B}.
\end{equation}

\subsubsection{Conservation of energy}

The total energy density can be written as a sum of internal energy $u$, kinetic energy, and electromagnetic field energy 
\begin{equation}
\label{eq:e-density}
    e_v' = \rho u + \frac{1}{2} \rho \vq{V}^2 + \frac{1}{2}\epsilon_0 \vq{E}^2 + \frac{1}{2 \mu_0} \vq{B}^2.
\end{equation}
However, it should be noted that with this definition, the kinetic energy of diffusion $\sum_\alpha \vq{d}_\alpha^2 / 2\rho_\alpha$ is ignored. This is justified because MHD equations apply only to systems near equilibrium, where all irreversible fluxes are small. In a single fluid model, the diffusive kinetic energy is a diffusive flux squared, so it is second order and may  be dropped.

In accordance with the energy density definition, the total energy flux is
\begin{equation}
\label{eq:heat-flux}
    \bm{\mathcal{J}}_e =  \rho u \vq{V} + \frac{1}{2} \rho \vq{V}^2 \vq{V} + \frac{1}{\mu_0} \vq{E} \times \vq{B} + \vq{P} \cdot \vq{V} + \vq{q},
\end{equation}
which consists of convection of internal energy, convection of plasma kinetic energy, convection of electromagnetic field energy, energy flux due to mechanical work, and heat flux. 
Equation~(\ref{eq:heat-flux}) serves to defines the heat flux $\vq{q}$. 
Defined in this way, $\vq{q}$ is not a true heat flux in the sense that it carries a contribution from diffusion. When species diffuse relative to one another, they carry thermal energy with them, contributing a factor of $\sum_{\alpha}h_\alpha \vq{d}_\alpha$ to the energy flux, \cite{bearman_1958,degroot_mazur} where $h_\alpha$ is the partial specific enthalpy of species $\alpha$. This contribution to the energy should not be confused with the kinetic energy of diffusion. The enthalpy diffusion term represents the diffusion of thermal energy, while the kinetic energy of diffusion represents the energy associated with the flow of species $\alpha$ with respect to the center of mass velocity. The former is a first order term and the latter is second order. 

Whether or not to include the enthalpy diffusion term in the definition of $\vq{q}$, or to separate it out, is a matter of convention. 
Most references include it in the definition of $\vq{q}$, including the Chapman-Enskog solution that is the basis for most descriptions of MHD in plasmas~\cite{chapman_cowling}. Additionally, this definition is easily computed from individual particle trajectories, which is important for closing the MHD equations through Green-Kubo relations. 
Thus, we choose the same convention here.

Definitions of energy density and heat flux from Eqs.~(\ref{eq:e-density}) and (\ref{eq:heat-flux}) may be inserted into the general conservation equation from Eq.~(\ref{eq:general-conservation}) to obtain the conservation of energy relation
\begin{equation}
    \label{eq:final-energy}
    \rho \frac{du}{dt} = - \nabla \cdot \vq{q} - \vq{P}:\nabla \vq{V} + \vq{j} \cdot \vq{E}',
\end{equation}
where $\vq{E}' = \vq{E} + \vq{V} \times \vq{B}$ is the electric field in the fluid reference frame.

\subsection{Electromagnetic field}
Conservation of mass, momentum, and energy
depend on the electromagnetic field, and therefore couple to Maxwell's equations. 
Since the quasineutrality assumption is made, Gauss's law does not influence MHD. 
Furthermore, a standard approximation in MHD is that the plasma and fields are non-relativistic, therefore excluding light waves and the displacement current in the Maxwell-Amp\'{e}re law. 
This leaves the standard ``pre-Maxwell'' equations 
\begin{subequations}
\label{eq:maxwells}
\begin{align}
\label{eq:maxa}
    \nabla \times \vq{E} &= - \frac{\partial \vq{B}}{\partial t} \\
    \label{eq:maxb}
    \nabla \times \vq{B} &= \mu_0 \vq{j}, \\
    \label{eq:maxc}
    \nabla \cdot \vq{B} &= 0 
\end{align}
\end{subequations}
to describe the electromagnetic field evolution.

Conservation of mass [Eq.~(\ref{eq:final-mass})], momentum [Eq.~(\ref{eq:final-mom})], and energy [Eq.~(\ref{eq:final-energy})], along with Maxwell's equations [Eq.~(\ref{eq:maxwells})] provide evolution equations for the fluid variables ($\rho, \vq{V}, u$), and fields ($\vq{E},\vq{B}$). 
However, they depend on the electrical current ($\vq{j}$), heat flux ($\vq{q}$) and pressure ($\vq{P}$), which must be related in some way to the fluid variables and fields to close the equations. In the next section, an expression for entropy production is developed and used to establish linear constitutive relations which satisfy this requirement. 

\subsection{Entropy production}
Many of the arguments in this section follow closely to those outlined in the standard reference on non-equilibrium thermodynamics by de Groot and Mazur~\cite{degroot_mazur}. It should be acknowledged that electromagnetic phenomena are treated in this text and other works in literature,~\cite{Maugin_1993,Wolff_1979,Kluitenberg_1973} but as far as the authors are aware, this tutorial presents the first complete derivation of closed MHD equations using non-equilibrium thermodynamics. 

It is an empirical fact that for a wide range of systems, thermodynamic fluxes are linear functions of thermodynamic forces when a system is near equilibrium. We can therefore write, in the most general case, that
\begin{equation}
\label{eq:general-linear}
    \bm{\mathcal{J}}_i = \sum_k L_{ik} \bm{X}_k
\end{equation}
where $\bm{\mathcal{J}}_i$ is any thermodynamic flux, $\bm{X}_k$ is thermodynamic force $k$, and the quantities $L_{ik}$ are \textit{phenomenological coefficients} that connect fluxes  to forces. Hence, fluxes may in principle be functions of \textit{all} thermodynamic forces in a system. When $i = k$, we call $\bm{\mathcal{J}}_i$ and $\bm{X}_k$ a conjugate force/flux pair. 

For Eq.~(\ref{eq:general-linear}) to be useful, a method for obtaining the forces and fluxes present in a system must be developed. 
Any relaxation process will produce entropy. 
It is possible to show that the entropy production per unit volume and time ($\sigma_S$) is a bilinear expression of the conjugate forces and fluxes present in a system~\cite{degroot_mazur}
\begin{equation}
    \sigma_S = \sum_i \bm{\mathcal{J}}_i \cdot \vq{X}_i.
\end{equation}
A suitable expression for $\sigma_S$ allows determination of the relevant thermodynamic forces and fluxes. To obtain such an expression, start by splitting the variation in total entropy $S$ into two terms
\begin{equation}
    \label{eq:entropy-var}
    dS = d_i S + d_e S,
\end{equation}
where $d_i S$ represents the entropy produced by the system (internal) and $d_e S$ represents the entropy supplied to the system by its surroundings (external). The second law of thermodynamics states $d_i S \geq 0$.

Defining $s$ to be the entropy per unit mass, 
\begin{equation}
\label{eq:S}
    S = \int_V \rho s d\mathcal{V}.
\end{equation}
Similarly, defining the total entropy flux $\bm{\mathcal{J}}_{S,\textrm{tot}}$ and using $\sigma_S$ defined above, 
\begin{equation}
\label{eq:disdes}
    \frac{d_i S}{dt} = \int_V \sigma_S d\mathcal{V} \quad \mathrm{and} \quad \frac{d_e S}{dt} = - \int_\Omega \bm{\mathcal{J}}_{S,\textrm{tot}} \cdot d \vq{\Omega}.
\end{equation}
The time derivative of Eq.~(\ref{eq:entropy-var}) may be taken and written in terms of the quantities in Eqs.~(\ref{eq:S}) and (\ref{eq:disdes}) to obtain
\begin{equation}
    \int_\mathcal{V} \left(\frac{\partial \rho s}{\partial t} + \nabla \cdot \bm{\mathcal{J}}_{S,\textrm{tot}} - \sigma_S \right) d\mathcal{V} = 0.
\end{equation}
This equation must hold for any volume $\mathcal{V}$, so 
\begin{equation}
    \frac{\partial \rho s}{\partial t} = - \nabla \cdot \bm{\mathcal{J}}_{S,\textrm{tot}} + \sigma_S.
\end{equation}
Alternatively, defining a diffusive entropy flux
\begin{equation}
    \bm{\mathcal{J}}_S = \bm{\mathcal{J}}_{S,\textrm{tot}} - \rho s \vq{V}
\end{equation}
the entropy balance equation can be written 
\begin{equation}
\label{eq:S_balance}
    \rho \frac{d s}{dt} = -\nabla \cdot \bm{\mathcal{J}}_S + \sigma_S.
\end{equation}

To obtain an expression for $\sigma_S$ in terms of the relevant fluid variables, first recall the specific Gibbs relation
\begin{equation}
    T ds = du + p d(1/\rho) - \sum_\alpha \mu_\alpha c_\alpha
\end{equation}
where $T$ is the temperature, $p$ is the pressure, and $\mu_\alpha$ is the partial specific chemical potential of species $\alpha$.
The assumption of local thermodynamic equilibrium will now be invoked. Though the system as a whole need not be in equilibrium, it is assumed that the fluid's domain may be broken into small elements which each satisfy the same equation of state as they would in a true global equilibrium. With this assumption, a time derivative of the Gibbs relation may be taken
\begin{equation}
\label{eq:dt-gibbs}
    T \frac{ds}{dt} = \frac{du}{dt} + p \frac{d (1/\rho)}{dt} - \sum_\alpha \mu_\alpha \frac{d c_\alpha}{dt}.
\end{equation}
Notably, expressions for the time derivatives appearing here were already obtained in Eqs.~(\ref{eq:final-mass}), (\ref{eq:partial-mass}), and (\ref{eq:final-energy}). Inserting these into Eq.~(\ref{eq:dt-gibbs}) and grouping terms yields
\begin{align}
\label{eq:S_rate}
    \nonumber \rho \frac{ds}{dt} &= - \nabla \cdot \left[ \frac{1}{T} \left( \vq{q} - \sum_\alpha \mu_\alpha \vq{d}_\alpha \right) \right] - \frac{1}{T^2} \vq{q} \cdot \nabla T  \\ 
    &- \sum_\alpha \vq{d}_\alpha \cdot \nabla \frac{\mu_\alpha}{T} + \frac{1}{T}\vq{j} \cdot \vq{E}' - \frac{1}{T} \hat{\vq{P}}: \nabla \vq{V}
\end{align}
where 
\begin{equation}
    \hat{\vq{P}} = \vq{P} - p\mathcal{I}
\end{equation}
is the viscous pressure tensor. Comparing Eqs.~(\ref{eq:S_balance}) and (\ref{eq:S_rate}), the diffusive entropy flux is identified as
\begin{equation}
\label{eq:J_S}
    \bm{\mathcal{J}}_S =  \frac{1}{T} \left(\vq{q} - \sum_\alpha \mu_\alpha \vq{d}_\alpha \right)
\end{equation}
and the entropy production as
\begin{equation}
\label{eq:sigma-step}
    \sigma_S = - \frac{1}{T^2} \vq{q} \cdot \nabla T - \sum_\alpha \vq{d}_\alpha \cdot \nabla \frac{\mu_\alpha}{T} + \frac{1}{T}\vq{j} \cdot \vq{E}' - \frac{1}{T} \hat{\vq{P}}: \nabla \vq{V}.
\end{equation}
The grouping of terms to identify $\bm{\mathcal{J}}_S$ and $\sigma_S$ may seem somewhat arbitrary. However, the separation is determined uniquely so long as $\sigma_S = 0$ at equilibrium and is invariant under Galilean transform,~\cite{degroot_mazur} both of which apply here. 

The sum over species in Eqs.~(\ref{eq:J_S}) and (\ref{eq:sigma-step}) may seem to indicate that a single fluid limit has not yet been obtained. 
However, this can be simplified by expanding in the small electron-to-ion mass ratio. 
Denoting electron values with subscript $e$, it is expected that $\vq{d}_e \gg \sum_i \vq{d}_i$, where $i$ indexes ion species, so that $\vq{j} \approx z_e \vq{d}_e$. Moreover, since the equations are based on the specific chemical potential (per unit mass), it is expected that $\mu_e \gg \sum \mu_i$. These two facts enable one to combine the diffusion and current density terms in Eq.~(\ref{eq:sigma-step}), so that
\begin{equation}
\label{eq:entropy-prod}
    \sigma_S = - \frac{1}{T^2} \vq{q} \cdot \nabla T + \vq{j} \cdot \left[\frac{\vq{E}'}{T} - \frac{1}{z_e} \nabla \frac{\mu_e}{T} \right] - \frac{1}{T} \hat{\vq{P}}: \nabla \vq{V}.
\end{equation}
This is the form of entropy production that will be used.  We may identify that heat flux is conjugate to a temperature gradient, electrical current is conjugate to an electric field and electron chemical potential gradient, and viscous momentum flux is conjugate to a velocity gradient. Linear constitutive relations may be written in terms of these forces and fluxes. 

\subsection{Linear constitutive relations}
As stated in Eq.~(\ref{eq:general-linear}), thermodynamic fluxes are, in principle, linear functions of \textit{all} thermodynamic forces. Hence, a general form of the constitutive relations that includes all thermodynamic forces identified in Eq.~(\ref{eq:entropy-prod}) has the form
\begin{subequations}
\label{eq:gen-fluxes}
\begin{align}
\label{eq:gen-current}
    \vq{j} &= \vq{L}_{jj} \cdot \left[\frac{\vq{E}'}{T} - \frac{1}{z_e} \nabla \frac{\mu_e}{T} \right]  - \vq{L}_{jq} \cdot \frac{\nabla T}{T^2} - \vq{L}_{jP} : \frac{\nabla \vq{V}}{T}, \\
\label{eq:gen-heat}
    \vq{q} &= \vq{L}_{qj} \cdot \left[\frac{\vq{E}'}{T} - \frac{1}{z_e} \nabla \frac{\mu_e}{T} \right]  - \vq{L}_{qq} \cdot \frac{\nabla T}{T^2} - \vq{L}_{qP} : \frac{\nabla \vq{V}}{T}, \\
    \hat{\vq{P}} &= \vq{L}_{Pj} \cdot \left[\frac{\vq{E}'}{T} - \frac{1}{z_e} \nabla \frac{\mu_e}{T} \right]  - \vq{L}_{Pq} \cdot \frac{\nabla T}{T^2} - \vq{L}_{PP} : \frac{\nabla \vq{V}}{T},
\end{align}
\end{subequations}
where the $\vq{L}$ terms are tensors of phenomenological coefficients.
Here the subscript notation for $\vq{L}_{lk}$ is defined such that $l$ refers to the flux in the relationship being considered, and $k$ refers to the flux that is conjugate to the particular force term that $\vq{L}_{lk}$ is multiplying. 
Hence, $\vq{L}_{jj}$ is a second-rank tensor, $\vq{L}_{jP}$ a third-rank tensor, and $\vq{L}_{PP}$ a fourth-rank tensor. 

In this section, symmetry arguments will be used to show that the vector forces and fluxes appearing in Eq.~(\ref{eq:entropy-prod}) do not couple with the tensor forces and fluxes. This is often referred to as Curie's principle~\cite{curie_1908} and sets all the third-rank tensor terms ($\vq{L}_{jP}$, $\vq{L}_{qP}$, $\vq{L}_{Pj}$, and $\vq{L}_{Pq}$) to zero, providing a considerable simplification. Additionally, the Onsager relations are presented,~\cite{Onsager_1931} which allow one to relate different phenomenological coefficients to each other.

\subsubsection{Orthogonal transformations \label{sec-trans}}
The arguments here are a generalization of what is provided in the reference by de Groot and Mazur~\cite{degroot_mazur} to include a magnetic field. 
 The forces and fluxes appearing in Eq.~(\ref{eq:entropy-prod}) are polar, meaning they change sign under reversal of the coordinate axes. This is perhaps not obvious due to the presence of a magnetic field, which appears in the term $\vq{V} \times \vq{B}$ in the definition of $\vq{E}'$. However, the cross product of a polar vector ($\vq{V}$) and an axial vector ($\vq{B}$) results in a polar vector.~\cite{arfken12} A consequence of the forces and fluxes being polar is that the phenomenological coefficients connecting vector fluxes to tensor forces (and vice versa) can be represented by a third rank polar tensor, denoted $L_{ijk}$. Recall that given some orthogonal transformation $\vq{A}$, a third-rank polar tensor transforms as
\begin{equation}
\label{eq:rank3}
   {L}'_{ijk} =\sum_{l,m,n} A_{il}A_{jm}A_{kn}  {L}_{lmn}
\end{equation}
where $L'_{ijk}$ is the transformed tensor. 

Let us now adopt a Cartesian coordinate system and take the magnetic field $\vq{B}$ to be in the $\hat{\vq{z}}$ direction. Note that this is not a restrictive assumption, because even if $\vq{B}$ is non-uniform in space, one could temporarily align each fluid element's z-axis with the magnetic field then transform to the desired orientation. Moreover, within each fluid element, $\vq{B}$ must be approximately constant, else the condition of a local thermodynamic equilibrium is not met. Inspection of the entropy production in Eq.~(\ref{eq:entropy-prod}) shows that the fluid is invariant under the following transformation
\begin{equation}
    \vq{A}_1 = 
    \begin{pmatrix}
        -1 & 0 & 0\\
        0 & -1 & 0\\
        0 & 0 &-1
    \end{pmatrix},
\end{equation}
i.e., the fluid obeys inversion symmetry. This directly follows from the fact that the entropy production is a bilinear expression of polar vectors. For the symmetry to be obeyed, one requires $L'_{ijk} = L_{ijk}$. However, when $\vq{A}_1$ is applied to Eq.~(\ref{eq:rank3}), one obtains $L'_{ijk} = -L_{ijk}$. It may therefore be concluded that $L_{ijk} = 0$, so the vector forces and fluxes do not couple with those of tensors in the linear constitutive relations. Physically, this means that heat fluxes and electrical currents may not arise from velocity gradients ($\vq{L}_{jP} = \vq{L}_{qP}=0$), and a viscous momentum flux may not arise from any of the vector forces ($\vq{L}_{Pj} = \vq{L}_{Pq}=0$).

Consider now the effect of an orthogonal transformation on the second rank tensors, which transform as
\begin{equation}
    L'_{ik} = \sum_{l,m} A_{il}A_{km} L_{lm}.
\end{equation}
In addition to inversion symmetry, each fluid element obeys a rotational symmetry about the z-axis. An infinitesimally small rotation about this axis is governed by the transformation matrix
\begin{equation}
\label{eq:rot-trans}
    \vq{A}_2 = 
    \begin{pmatrix}
        1 & -\alpha & 0\\
        \alpha & 1 & 0\\
        0 & 0 & 1
    \end{pmatrix}
\end{equation}
where $\alpha$ is the small angle. Consistent with the infinitesimal rotation, any term of order $\alpha^2$ that appears in the transformed matrix is dropped. The requirement that $L'_{ik} = L_{ik}$ implies $L_{xy} = - L_{yx}$, $L_{xx} = L_{yy}$, and all terms with one z-index are zero. A general rank two tensor in this system can therefore be written
\begin{equation}
    \vq{L} = 
    \begin{pmatrix}
        L_{xx} & L_{xy} & 0\\
        -L_{xy} & L_{xx} & 0\\
        0 & 0 & L_{zz}
    \end{pmatrix},
\end{equation}
with just three independent coefficients. 
This applies to the $\vq{L}_{jj}, \vq{L}_{jq}, \vq{L}_{qj}$, and $\vq{L}_{qq}$ terms in Eq.~(\ref{eq:gen-fluxes}). 

It is common in plasma physics to define a coordinate system aligned with the magnetic field. For any vector $\vq{r}$, the coordinate system is defined such that
\begin{subequations}
\begin{eqnarray}
    \vq{r}_{\parallel} &=& \hat{\vq{b}}(\hat{\vq{b}} \cdot \vq{r}) \\
    \vq{r}_{\perp} &=& \vq{r} - \vq{r}_{\parallel} \\
    \vq{r}_{\wedge} &=& \hat{\vq{b}} \times \vq{r} ,
\end{eqnarray}
\end{subequations}
where $\hat{\vq{b}}$ is the direction of the magnetic field. With this, a general rank-two tensor of phenomenological coefficients may be written as
\begin{equation}
\label{eq:gen-rank2}
    \vq{L} = 
    \begin{pmatrix}
        L_{\perp} & -L_{\wedge} & 0\\
        L_{\wedge} & L_{\perp} & 0\\
        0 & 0 & L_{\parallel}
    \end{pmatrix}.
\end{equation}
In this form, the dot product of a second-rank tensor of coefficients with an arbitrary vector force $\vq{X}$ is 
\begin{equation}
    \vq{L} \cdot \vq{X} = L_{\parallel} {X}_\parallel + L_{\perp} {X}_\perp + L_{\wedge} {X}_\wedge, 
\end{equation}
which define the parallel, perpendicular, and cross directions. 

\subsubsection{Onsager-Casimir relations}
In addition to spatial symmetries, time reversal symmetry may be used to simplify the linear constitutive relations. The Onsager relations are derived from this symmetry~\cite{Onsager_1931}, and their generalization to a magnetized system is~\cite{Casimir_1945} 
\begin{equation}
\label{eq:onsager}
    \vq{L}_{\mu \nu} (\vq{B}) = \vq{L}_{\nu \mu}^\top (-\vq{B})
\end{equation}
where $\mu$ and $\nu$ index the different forces and fluxes and $\top$ represents the transpose. The sign flip occurs because a system with a magnetic field satisfies time reversal symmetry only if the field is also reversed. This modification was found by Casimir~\cite{Casimir_1945}, and thus the class of relations is sometimes called the Onsager-Casimir relations. This relates coefficients both within and across different tensors to each other, again allowing one to reduce the number of independent coefficients.

If the Onsager-Casimir relations are applied to $\vq{L}$ in Eq.~(\ref{eq:gen-rank2}), one can see that the off-diagonal components must flip signs upon reversal of $\vq{B}$
\begin{equation}
\label{eq:xy-prop}
    L_{\wedge} (\vq{B}) = -L_{\wedge}(-\vq{B}). 
\end{equation}
but the diagonal components are unchanged
\begin{equation}
\label{eq:xx-prop}
    L_{\perp} (\vq{B}) = L_{\perp}(-\vq{B}) \quad \mathrm{and} \quad L_{\parallel} (\vq{B}) = L_{\parallel}(-\vq{B}). 
\end{equation}
These properties will prove useful in the next section.

\subsubsection{Current density (Ohm's Law) \label{sec:ohms}}
It has been shown that $\vq{L}_{jP} = 0$, so the expression for current density from Eq.~(\ref{eq:gen-current}) can be written
\begin{equation}
\label{eq:lin-elec}
    \vq{j} = \vq{L}_{jj} \cdot \left[\frac{\vq{E}'}{T} - \frac{1}{z_e} \nabla \frac{\mu_e}{T} \right]  - \vq{L}_{jq} \cdot \frac{\nabla T}{T^2}
\end{equation}
where phenomenological coefficients $ \vq{L}_{jj}$ and $\vq{L}_{jq}$ are second rank tensors. It is common to define the electrical conductivity $\bm{\sigma} = \vq{L}_{jj}/T$ and electrothermal coefficient $\bm{\varphi} = -\vq{L}_{jq}/T^2$, so that
\begin{equation}
\label{eq:current-final}
    \vq{j} = \bm{\sigma} \cdot \left[\vq{E}' - \frac{T}{z_e} \nabla \frac{\mu_e}{T} \right] + \bm{\varphi} \cdot \nabla T.
\end{equation}
This is a generalized Ohm's law for plasmas with a non-ideal equation of state. Notably, it depends on the electron chemical potential gradient, which is a non-trivial calculation beyond the ideal gas approximation. Further discussion on this is provided in Sec.~\ref{sec:eos}. 

Using the symmetry arguments presented in the previous section, the electrical conductivity tensor and electrothermal tensor are composed of three independent coefficients each. The electrical conductivity is therefore
\begin{equation}
    \bm{\sigma} = 
    \begin{pmatrix}
        \sigma_{\perp} & -\sigma_{\wedge} & 0\\
        \sigma_{\wedge} & \sigma_{\perp} & 0\\
        0 & 0 & \sigma_{\parallel}
    \end{pmatrix}.
\end{equation}
It is seen that an electric field in the y-direction yields a current in the x-direction, and vice versa. This is the Hall effect, and it should be stressed here that even within a one fluid model, it can be captured so long as one defines a proper electrical conductivity tensor. 
Similarly, the electrothermal tensor is
\begin{equation}
    \bm{\varphi} = 
    \begin{pmatrix}
        \varphi_{\perp} & -\varphi_{\wedge} & 0\\
        \varphi_{\wedge} & \varphi_{\perp} & 0\\
        0 & 0 & \varphi_{\parallel}
    \end{pmatrix}.
\end{equation}
Ohm's law is seen to have six independent coefficients.

\subsubsection{Heat flux (Fourier's law)}
The heat flux equation may be obtained in similar fashion to that of the current density. Since $\vq{L}_{qP} = 0$, the heat flux from Eq.~(\ref{eq:gen-heat}) is 
\begin{equation}
    \vq{q} = - \vq{L}_{qq} \cdot \frac{\nabla T}{T^2} + \vq{L}_{qj} \cdot \left[\frac{\vq{E}'}{T} - \frac{1}{z_e} \nabla \frac{\mu_e}{T} \right],
\end{equation}
where $\vq{L}_{qj}$ and $\vq{L}_{qq}$ are second rank tensors of phenomenological coefficients. Defining a thermal conductivity $\bm{\lambda} = \vq{L}_{qq}/T^2$ and thermoelectric coefficient $\bm{\phi} = \vq{L}_{qj}/T$, the heat flux is
\begin{equation}
\label{eq:heat-final}
    \vq{q} = -\bm{\lambda} \cdot \nabla T + \bm{\phi} \cdot \left[\vq{E}' - \frac{T}{z_e} \nabla \frac{\mu_e}{T} \right].
\end{equation}
Following the symmetry arguments of the previous section, the thermal conductivity is composed of three independent coefficients
\begin{equation}
    \bm{\lambda} = 
    \begin{pmatrix}
        \lambda_{\perp} & -\lambda_{\wedge} & 0\\
        \lambda_{\wedge} & \lambda_{\perp} & 0\\
        0 & 0 & \lambda_{\parallel}
    \end{pmatrix}.
\end{equation}

Because they are based on the same non-conjugate force-flux pairs, components of the thermoelectric tensor ($\bm{\phi}$) can be written entirely in terms of the electrothermal coefficients ($\bm{\varphi}$). It follows from the Onsager-Casimir relations, Eq.~(\ref{eq:onsager}), the definition of the electrothermal coefficient, $\bm{\varphi} = -\vq{L}_{jq}/T^2$, and the definition of the thermoelectric coefficient, $\bm{\phi} = \vq{L}_{qj}/T$, that 
\begin{equation}
    \bm{\phi}(\vq{B}) = - \bm{\varphi}^\top (-\vq{B}) T.
\end{equation}
Applying the properties from Eqs.~(\ref{eq:xy-prop}) and (\ref{eq:xx-prop}) the components of $\bm{\phi}$ can be written in terms of those in $\bm{\varphi}$ as
\begin{equation}
    \bm{\phi} = T\begin{pmatrix}
        -\varphi_{\perp} & \varphi_{\wedge} & 0\\
        -\varphi_{\wedge} & -\varphi_{\perp} & 0\\
        0 & 0 & -\varphi_{\parallel}
    \end{pmatrix}.
\end{equation}
Due to the Onsager symmetry, Ohm's law and Fourier's law are together composed of nine independent coefficients.

\subsubsection{Viscosity (Newton's law)}
Since all third rank tensors have been shown to be zero, the linear constitutive relation for viscosity is
\begin{equation}
\label{eq:final-visc}
    \hat{\vq{P}} = - \vq{L}_{PP} : \frac{\nabla \vq{V}}{T},
\end{equation}
where $\vq{L}_{PP}$ is a fourth rank tensor, nominally composed of 81 coefficients. Symmetries reduce this to just seven independent coefficients. 
First, the pressure tensor is symmetric so long as the fluid's constituent particles posses no net angular momentum. To show this, let the total angular momentum density $\vq{l}$ may be written
\begin{equation}
    \vq{l} = \vq{r} \times \left(\rho\vq{V} + \epsilon_0 \vq{E} \times \vq{B} \right).
\end{equation}
Taking a time derivative and substituting the momentum equation from Eq.~(\ref{eq:mom-start}) yields
\begin{equation}
    \frac{\partial \vq{l}}{\partial t} = -\vq{r} \times \left[\nabla \cdot \left( \rho \vq{V} \vq{V} + \vq{P} - \bm{\mathcal{T}}_M \right)  \right].
\end{equation}
Algebraic manipulation allows one to write this in terms of the divergence of a flux and source term as
\begin{equation}
    \frac{\partial \vq{l}}{\partial t} = - \nabla \cdot \left[\vq{r} \times \left( \rho \vq{V} \vq{V} + \vq{P} - \bm{\mathcal{T}}_M \right)  \right] + \frac{1}{2} \vq{P}^{(a)},
\end{equation}
where $\vq{P}^{(a)}$ is the antisymmetric part of $\vq{P}$. Maxwell's stress tensor is symmetric, so its anti-symmetric component does not appear in the above equation. In accordance with the general form of conservation equations defined in Eq.~(\ref{eq:general-conservation}), it is clear that for momentum to be conserved, one must have $\vq{P}^{(a)} = 0$; i.e., the viscous pressure tensor must be symmetric.

Index notation will now be adopted. To avoid confusion, the tensor of coefficients $\vq{L}_{PP}$ will be written simply as $L$, so that subscripts now represent the index of the tensor. Using the fact that the viscous pressure tensor is symmetric, the phenomenological viscosity equation is
\begin{equation}
\label{eq:Pij}
    \hat{P}_{ij} = - L_{ijkl} \frac{(\nabla \vq{V})_{kl}^{(s)}}{T},
\end{equation}
where $(\nabla \vq{V})_{kl}^{(s)}$ is the symmetric part of the velocity gradient tensor. Due to symmetry, $\hat{P}_{ij}$ and $(\nabla \vq{V})_{kl}^{(s)}$ each contain just six independent coefficients, so a form of Eq.~(\ref{eq:Pij}) can be constructed that contains 36 independent phenomenological coefficients. Expressing this mathematically is aided by using Voigt notation. 

Let $1 = xx$, $2 = yy$, $3 = zz$, $4 = yz$, $5 = zx$, and $6 = xy$. The pressure tensor can be expressed as a vector $\hat{P}_i$ with six elements, indexed by the 1-6 defined here (e.g. $\hat{P}_4 = \hat{P}_{yz} = \hat{P}_{zy}$). For the sake of brevity, we will write the symmetric velocity gradient as $\vq{W} = (\nabla\vq{V})^{(s)}$, so that in Voigt notation one has $W_i$, where $W_1 = (\nabla \vq{V})_{xx}^{(s)}$ and $W_4 = 2(\nabla \vq{V})_{yz}^{(s)}$, etc. Similar expressions exist for the other components, and a factor of 2 appears when $i = $ 4, 5, and 6 because the terms with mixed indices appear twice in the linear constitutive equation. The coefficient of tensors can be written as a second rank ($6\times 6$) tensor $L_{ij}$ with 36 independent coefficients. With this notation, the linear constitutive relation is written in simpler form as
\begin{equation}
    \hat{P}_i = - \frac{1}{T}L_{ij} W_{j}.
\end{equation}

Symmetry arguments will now be used to further reduce the number of coefficients. Orthogonal transformations cannot be applied directly to the second rank tensor $L_{ij}$. Instead, they must be applied to the fourth rank tensor, and the results can be subsequently rewritten in terms of the second rank tensor. Applying the transformation associated with rotational symmetry from Eq.~(\ref{eq:rot-trans}), the second rank tensor can be reduced to 
\begin{equation}
    L_{ij} = 
    \begin{pmatrix}
        L_{11} & L_{12} & L_{13} & 0 & 0 & L_{16}\\
        L_{12} & L_{11} & L_{13} & 0 & 0 & -L_{16}\\
        L_{13} & L_{13} & L_{33} & 0 & 0 & 0 \\
        0 & 0 & 0 & L_{44} & L_{45} & 0 \\
        0 & 0 & 0 & -L_{45} & L_{44} & 0 \\
        -L_{16} & L_{16} & 0 & 0 & 0 & L_{66}
    \end{pmatrix}.
\end{equation}
The Onsager relations were also used here to equate $L_{13} = L_{31}$ and $L_{23} = L_{32}$. See Refs.~\onlinecite{hooyman_1954} and \onlinecite{Scheiner_2020} for details. 

It is now seen that there are seven independent viscosity coefficients. It is common to distinguish between shearing and dilatational motion by splitting the symmetric velocity gradient into its trace and traceless parts. The traceless part is simply the rate-of-shear tensor $\vq{S}$, given by
\begin{equation}
  \vq{S} =  \frac{1}{2} \bigl[ \nabla \vq{V} + (\nabla \vq{V})^\top - \frac{2}{3}  (\nabla \cdot \vq{V})\vq{I}],
\end{equation}
and the trace will be written explicitly as $\frac{1}{3}(\nabla \cdot \vq{V}) \vq{I}$. In Voigt notation, $S_i = W_i - \nabla \cdot \vq{V} / 3$ for $i = 1$ to $3$. The viscous pressure tensor must similarly be split $\hat{\vq{P}}^{\circ}= \hat{\vq{P}} - \mathrm{Tr}(\hat{\vq{P}})(\vq{I}/3)$. To connect with the common notation, define the following coefficients:
\begin{subequations}
\begin{eqnarray}
    \eta_0 &=& \frac{1}{6T} (2 L_{33} - 4L_{13} + L_{11} + L_{12}) \\
    \eta_1 &=& \frac{1}{6T} (2L_{11} - L_{12} - 2L_{13}+L_{33}) \\
    \eta_2 &=& \frac{1}{T}L_{44} \\
    \eta_3 &=& \frac{1}{T}L_{16} \\
    \eta_4 &=& \frac{1}{T}L_{45} \\
    \eta_v &=& \frac{1}{9T} (2L_{11} + 2L_{12} + 4L_{13} + L_{33}) \\
    \zeta &=& \frac{1}{3T} ( L_{13} + L_{33} - L_{11} - L_{12}),
\end{eqnarray}
\end{subequations}
leading to the following alternative form of Eq.~(\ref{eq:Pij}) 
\begin{widetext}
\begin{equation}
    \begin{blockarray}{cccccccc}
    & s_1 & s_2 & s_3 & w_4 & w_5 & w_6 & \nabla \cdot \vq{V}/3 \\
    \begin{block}{c(ccccccc)}
        \hat{P}^{\circ}_{1} & -2 \eta_1 & 2(\eta_1 - \eta_0) & 0 & 0 & 0 & -\eta_3 & \zeta \\
        \hat{P}^{\circ}_{2} & 2(\eta_1 - \eta_0) & -2\eta_1 & 0 & 0 & 0 & \eta_3 & \zeta \\
        \hat{P}^{\circ}_{3} & 0 & 0 & -2 \eta_0 & 0 & 0 & 0 & -2\zeta \\
        \hat{P}_{4} & 0 & 0 & 0 & -\eta_2 & -\eta_4 & 0 & 0 \\
        \hat{P}_{5} & 0 & 0 & 0 & \eta_4 & -\eta_2 & 0 & 0 \\
        \hat{P}_{6} & \eta_3 & - \eta_3 & 0 & 0 & 0 & \eta_0 - 2\eta_1 & 0 \\
        \mathrm{Tr}\hat{P} & \zeta & \zeta & -2\zeta & 0 & 0 & 0 & -9 \eta_v \\
    \end{block}
    \end{blockarray}
\end{equation}
\end{widetext}
It is seen that $\eta_1, \eta_2, \eta_3, \eta_4$ and $\eta_5$ can be considered coefficients of shear viscosity, $\eta_v$ the coefficient of bulk viscosity, and $\zeta$ a cross-term. The form presented here resembles what is common in neutral fluid dynamics~\cite{degroot_mazur}. 

In plasmas, it is more common for viscosity coefficients to be written as in the Braginskii review~\cite{braginskii_review}. The review makes use of the Chapman-Enskog procedure, which due to the assumptions of an ideal monatomic gas, implies that the bulk viscosity $\eta_v$ and cross-coefficient $\zeta$ are zero. This is generally a good assumption in classical atomic plasmas, even at high densities~\cite{LeVan_2025}. However, bulk viscosity can be large in the presence of electronic excitation and molecular degrees of freedom~\cite{Istomin_2017, LeVan_2024}, so here we generalize these expressions to account for the bulk and cross viscosity terms. 

To connect with the Braginskii form, define the following coefficients~\cite{Scheiner_2020}
\begin{subequations}
    \begin{eqnarray}
        \eta_0^B &=& \eta_0 \\
        \eta_1^B &=& 2 \eta_1 - \eta_0 \\
        \eta_2^B &=& \eta_2 \\
        \eta_3^B &=& \eta_3 \\
        \eta_4^B &=& -\eta_4, 
    \end{eqnarray}
\end{subequations}
with which the linear constitutive relation for viscosity may be written  
\begin{subequations}
\label{eq:newton}
    \begin{align}
    \label{eq:linpa}
        \hat{\vq{P}}^{\circ}_{xx} &=  - \eta_0^B (S_{xx} + S_{yy}) \\ 
        \nonumber &-  \eta_1^B (S_{xx} - S_{yy}) - 2 \eta_3^B S_{xy} +\frac{1}{3} \zeta \nabla \cdot \vc{V} \\
        \hat{\vq{P}}^{\circ}_{yy} &=  - \eta_0^B (S_{xx} + S_{yy}) \\ 
        \nonumber &+  \eta_1^B (S_{xx} - S_{yy}) + 2 \eta_3^B S_{xy} +\frac{1}{3} \zeta \nabla \cdot \vc{V} \\
        \hat{\vq{P}}^{\circ}_{zz} &= -2\eta_o^B S_{zz} - \frac{2}{3} \zeta \nabla \cdot \vq{V}\\
        \hat{\vq{P}}_{xy} &= - 2\eta_1^B S_{xy} + \eta_3^B (S_{xx} - S_{yy}) \\
        \hat{\vq{P}}_{xz} &= - 2\eta_2^B S_{xz} - 2\eta_4^B S_{yz} \\
         \label{eq:linpf}\hat{\vq{P}}_{yz} &= -2 \eta_2^B S_{yz} + 2\eta_4^B S_{xz} \\
         - \mathrm{Tr}(\hat{\vq{P}})/3 &= \zeta S_{zz} + \eta_v \nabla \cdot \vq{V} .
    \end{align}
\end{subequations}
The property $S_{xx} + S_{yy} + S_{zz}=0$ was used in obtaining this result. Note that the viscous pressure tensor is related to the symmetric pressure tensor by 
\begin{equation}
\hat{\vq{P}} = \hat{\vq{P}}^{\circ} - (\zeta \vq{S}\colon \hat{\vq{b}} \hat{\vq{b}} + \eta_v \nabla \cdot \vq{V}) \vq{I} .
\end{equation}

An advantage of this formulation is that the viscous pressure tensor may be written in a coordinate system aligned with the magnetic field, as done with the previous transport coefficients. This can be seen by first defining the following traceless tensors: 
\begin{subequations}
    \begin{align}
        W'_{(1)} &=  2(\vq{I} - \vq{b} \vq{b}) \cdot \vq{S} \cdot (\vq{I - \vq{bb}}) \\ 
        \nonumber &- (\vq{I} - \vq{b} \vq{b})(\vq{I} - \vq{b}\vq{b}):\vq{S} \\
        W'_{(2)} &=  2 (\vq{I} - \vq{b}\vq{b}) \cdot \vq{S} \cdot \vq{b} \vq{b} \\ \nonumber &+ 2 \vq{b}\vq{b} \cdot
        \vq{S} \cdot (\vq{I} - \vq{b} \vq{b})\\ 
        W''_{(1)} &= 2\vq{b} \times \vq{S} \cdot (\vq{I} - \vq{b} \vq{b}) \\ \nonumber &- 2(\vq{I} - \vq{b}\vq{b})\cdot \vq{S} \times \vq{b}\\
         W''_{(2)} &= 2\vq{b} \times \vq{S} \cdot \vq{b} \vq{b} - 2 \vq{b} \vq{b} \cdot \vq{S} \times \vq{b} .
    \end{align}
\end{subequations}
Now applying $\hat{\vq{P}} = \hat{\vq{P}}_\parallel + \hat{\vq{P}}_\perp +\hat{\vq{P}}_\wedge$, the linear constitutive relations from Eqs.~(\ref{eq:linpa})-(\ref{eq:linpf}) may be rewritten as
\begin{subequations}
\label{eq:P_final}
    \begin{align}
        \hat{\vq{P}}^\circ_\parallel &= - \bigl[ 3 \eta_0^B (\vq{b} \cdot \vq{S} \cdot \vq{b}) - \zeta \nabla \cdot \vc{V} \bigr](\vq{b} \vq{b} - \frac{\vq{I}}{3}) \\
        \hat{\vq{P}}^\circ_\perp &= - \eta_1^BW'_{(1)} - \eta_2^B W'_{(2)} \\
        \hat{\vq{P}}^\circ_\wedge &= \frac{\eta_3^B}{2}W''_{(1)} + \eta_4^B W''_{(2)}.
    \end{align}
\end{subequations}
Physically, it is seen that $\eta_0^B$ describes stress parallel to the magnetic field, $\eta_1^B$ and $\eta_2^B$ perpendicular to the field, and $\eta_3^B$ and $\eta_4^B$ represent cross stresses.

\subsection{Discussion \label{sec:disc}}

Macroscopic considerations alone have now provided a set of MHD equations which are closed insofar as one knows the materials properties of the plasma. 
Conservation of mass [Eq.(\ref{eq:final-mass})], momentum [Eq.~(\ref{eq:final-mom})], and energy [Eq.~(\ref{eq:final-energy})], along with Maxwell's equations~(\ref{eq:maxwells}) provide evolution equations for the fluid variables ($\rho, \vq{V}, u$), and fields ($\vq{E},\vq{B}$). 
These depend on the unknown fluxes [electrical current ($\vq{j}$), heat flux ($\vq{q}$) and viscous pressure ($\hat{\vq{P}}$)] which are related to the forces that generate them [electric field ($\vq{E}'$), pressure gradient ($\nabla p$), chemical potential gradient ($\nabla  \mu_e$), temperature gradient ($\nabla T$) and velocity gradient ($\nabla \vq{V}$)] by the linear constitutive relations [Ohm's law [Eq.~(\ref{eq:current-final})], Fourier's law [Eq.~(\ref{eq:heat-final})], and Newton's law [Eq.~(\ref{eq:newton})]]. 
This constitutes a closed set of equations in terms of the materials properties of the plasma: electrical conductivity coefficients ($\bm{\sigma}$), electrothermal coefficients ($\bm{\varphi}$), thermal conductivity coefficients ($\bm{\lambda}$), and viscosity coefficients ($\bm{\eta}, \zeta$, and $\eta_v$), as well as an equation of state needed to relate the specific internal energy $u$, scalar pressure $p$, and specific electron chemical potential $\mu_e$ to the mass density $\rho$ and temperature $T$. 

A variety of ways can be envisioned to obtain these materials properties. 
Before there was knowledge of atoms as the constituents of matter, one had to rely on experimental measurements. 
With modern statistical physics, one may (in principle) relate them to statistical properties of microscopic dynamics of the constituent neutral atoms, ions and electrons. 
Section~\ref{sec:micro} discusses a general way to relate the materials properties of the plasma to correlations of fluctuations in the particle positions and velocities at equilibrium when the plasma can be described by classical statistical physics. 
Another common method is to develop a kinetic equation and solve it perturbatively for the MHD equations, such as the Boltzmann equation and Chapman-Enskog solution.~\cite{ferziger_kaper,chapman_cowling} 
This provides explicit formulas for the transport coefficients and equations of state, but is only valid in a dilute ideal-gas limit.

Before moving to the general procedure for relating macroscopic materials properties to microscale fluctuations, we note that in the classical limit thermodynamic quantities can be split into ``ideal'' and ``excess'' components~\cite{hansen_2013}
\begin{subequations}
\label{eq:eos_separation}
\begin{align}
    u = u_{\mathrm{ideal}} + u_{\mathrm{ex}} \\
    p = p_{\mathrm{ideal}} + p_{\mathrm{ex}} \\
    \mu_e = \mu_{e, \mathrm{ideal}} + \mu_{e,\mathrm{ex}}, 
\end{align}
\end{subequations}
where the ideal components have explicit expressions in terms of the state variables~\cite{Reif_1965}  
\begin{subequations}
\label{eq:ideal}
\begin{gather}
     \rho u_{\mathrm{ideal}} = \frac{3}{2} n k_B T \\
    p_{\mathrm{ideal}} =  n k_B T \\
    \mu_{e,\mathrm{ideal}} = -\frac{k_B T}{m_e} \ln \left[ \frac{m_e}{\rho_e} \left( \frac{2 \pi m_e k_B T}{h^2}\right)^{3/2}\right].\label{eq:mu_e_ideal}
\end{gather}
\end{subequations}
Here, $n=\sum_\alpha \rho_\alpha/m_\alpha$ is the number density, and $h$ is Planck's constant. It should also be noted that the thermodynamic quantities rely on the mass density of individual species. Hence for closure, one must know how to relate the total mass density $\rho$ to the mass density of a particular species $\rho_\alpha$. For two-component systems, quasi-neutrality ensures such a relation, but beyond this, one must be careful in applying the single fluid model. 

For systems that have quantum degenerate electrons, such as a dense plasma or liquid metal, the ideal components differ from Eq.~(\ref{eq:ideal}). 
For example, in the quantum free-electron gas, Eq.~(\ref{eq:ideal}) is replaced by~\cite{pathria2011,ichimaru_book}
\begin{subequations}
    \begin{align}
        \rho u_{\textrm{ideal}}^F &= \frac{3}{2} nk_BT \Theta^{3/2} I_{3/2}(\alpha) \\
        p_{\textrm{ideal}}^F &= nk_BT \Theta^{3/2} I_{3/2}(\alpha) \\
        I_{1/2}(\alpha) &= \frac{2}{3}\Theta^{-3/2}
    \end{align}
\end{subequations}
where the chemical potential enters through $\alpha \equiv \mu_e/k_BT$, the degeneracy parameter is $\Theta = k_BT/E_F$, the Fermi energy is $E_F = \hbar^2 (3\pi^2 n)^{2/3}/(2m_e)$, and the Fermi integral is defined as 
\begin{equation}
    I_\nu(\alpha) \equiv \int_0^\infty \frac{t^\nu}{\exp(t-\alpha) + 1} .
\end{equation}
This demonstrates the distinction between classical and quantum regimes even in an ideal limit. 

Considering a classical statistical description, the excess components of the equation of state can be related to correlations of the relative positions of particles at equilibrium. 
Similarly, the transport coefficients can be related to time correlations of the particle positions, velocities and relative forces at equilibrium. 
In principle, these correlation functions may be evaluated from experiment or computation. 
The next section develops this connection between the material properties and microscopic structure and dynamics.

\section{Microscopic considerations\label{sec:micro}}

\subsection{Irving-Kirkwood procedure}

Complimentary to the macroscopic arguments, one can derive conservation laws for the fluid scale by averaging the equations of motion of particles at the microscopic scale. 
Equating the results with the linear constitutive relations obtained from the macroscopic arguments shows the connection between transport coefficients and microscopic dynamics. 
Irving and Kirkwood pioneered a method to do this using a singlet density formulation in which local densities are related to Dirac delta functions~\cite{Irving_Kirkwood}. The procedure they developed results in expressions for thermodynamic fluxes in terms of ensemble averages over molecular distribution functions. The derivation will be briefly outlined here.

In the Gibbs's formulation of statistical mechanics, a classical system is characterized by the phase space probability density~\cite{hansen_2013,Evans_Morriss}
\begin{equation}
    f^{[N]}(\vq{r}^N, \vq{p}^N;t)
\end{equation}
where $\vq{r}^N = \vq{r}_1,..., \vq{r}_N$ and $\vq{p}^N = \vq{p}_1,..., \vq{p}_N$ are the positions and momenta of all particles in the system. This function is defined such that $f^{[N]} d\vq{r}^N d\vq{p}^N$ represents the probability that at a time $t$, the system lies in phase space element $d\vq{r}^N d\vq{p}^N$. The expectation value of variable $A(\vq{r}^N, \vq{p}^N)$ at time $t$ is given by an ensemble average
\begin{equation}
\label{eq:ensemble-avg}
    \Bigl \langle A(\vq{r}^N, \vq{p}^N) \Bigr \rangle = \int A f^{[N]}(\vq{r}^N, \vq{p}^N;t) d\vq{r}^N d\vq{p}^N,
\end{equation}
where the integral is over a fluid element scale that is infinitesimal at the macroscopic scale. 
Fluid variables can be defined as ensemble averages over microscopic particle properties. Here, the position vector $\vq{r}$ represents the location of some fluid element and $\vq{r}_i$ represents the location of particle $i$.
With this, the total mass density is defined to be
\begin{equation}
    \rho (\vq{r},t) = \Bigl \langle \sum_{i=1}^{N} m_i \delta (\vq{r} - \vq{r}_i) \Bigr \rangle,
\end{equation}
the kinetic momentum density as
\begin{equation}
    \rho(\vq{r}, t) \vq{V}(\vq{r}, t) = \Bigl \langle \sum_{i=1}^N m_i \vq{v}_i \delta(\vq{r} - \vq{r}_i)\Bigr \rangle,
\end{equation}
and energy density of matter as
\begin{equation}
    \rho(\vq{r}, t) e(\vq{r},t) =  \Bigl \langle \left[ \frac{1}{2} \sum_{i=1}^N m_i \vq{v}_i^2 + \frac{1}{2} \sum_{i,j}^N \phi_{ij}\right] \delta(\vq{r} - \vq{r}_i) \Bigr \rangle,
\end{equation}
where $\phi_{ij}$ is the interaction potential between particle $i$ and $j$. 

Time derivatives of these quantities provide conservation equations resembling those derived based on macroscopic arguments. These are made easier by exploiting the property found by Irving and Kirkwood~\cite{Irving_Kirkwood}
\begin{equation}
\label{eq:ik-theorem}
    \frac{\partial}{\partial t} \langle A\rangle = \sum_{i=1}^N\Bigl \langle \vq{v}_i \cdot \frac{\partial A}{\partial \vq{r}_i} + \vq{F}_i \cdot \frac{\partial A}{\partial \vq{p}_i} \Bigr \rangle,
\end{equation}
where the force on particle $i$ is
\begin{equation}
    \vq{F}_i = \sum_{j\neq i}^N \left(-\frac{\partial \phi_{ij}}{\partial \vq{r}_i}\right) + q_i \left(\vq{E} + \vq{v}_i \times \vq{B}\right).
\end{equation}
The first term results from microscopic interactions within a fluid element while the second term results from macroscopic forces. 

Considering  $A = \{\rho, \rho \vq{V}, \rho e\}$, Eq.~(\ref{eq:ik-theorem}) provides expressions for the conservation of mass, momentum and energy that are identical to Eqs.~(\ref{eq:final-mass}), (\ref{eq:final-mom}), and (\ref{eq:final-energy}), but with explicit expressions for the thermodynamic fluxes in terms of particle properties. 
Specifically, the current density is found to be 
\begin{equation}
    \vq{j} = \sum_{i=1}^N\Bigl \langle  q_i \vq{v}_i \delta(\vq{r} - \vq{r}_i) \Bigr \rangle,
\end{equation}
the pressure tensor
\begin{equation}
    \vq{P} = \sum_{i=1}^N\Bigl \langle \left[ m_i (\vq{v}_i - \vq{V})(\vq{v}_i - \vq{V}) + \frac{1}{2}\sum_{j \neq i}^N \vq{r}_{ij} \frac{\partial \phi_{ij}}{\partial \vq{r}_i} \right] \delta (\vq{r} - \vq{r}_i)\Bigr \rangle,
\end{equation}
and the heat flux
\begin{eqnarray}
    \nonumber \vq{q} = \sum_{i=1}^N\Bigl \langle  \Biggr[ (\vq{v}_i - \vq{V})\left(\frac{1}{2}m_i (\vq{v}_i - \vq{V})^2 + \frac{1}{2}\sum_{j \neq i}^N \phi_{ij} \right) \\
    + \frac{1}{2} \sum_{j \neq i}^N \vq{r}_i \cdot (\vq{v}_i - \vq{V})  \frac{\partial \phi_{ij}}{\partial \vq{r}_i}\Biggr] \delta(\vq{r} - \vq{r}_i)\Bigr \rangle.
\end{eqnarray}
In obtaining the above expressions, a uniform density has been assumed at the scale of the volume ($\mathcal{V}$) that defines a fluid element. 
This is simply related to the same condition of local thermodynamic equilibrium (LTE) that was assumed in the macroscopic thermodynamics arguments. 
The delta functions can be replaced with the appropriate volume factor $1/\mathcal{V}$, since all particles are considered to be within the fluid element at position $\vq{r}$. The center of mass velocity $\vq{V}$ may also be dropped if the center of mass position is chosen as the frame of reference, since it is zero at equilibrium in this reference frame. 
Furthermore, since the expressions were derived from conservation equations that must hold at any instant in time $t$, the average may be dropped so that instantaneous representations of the fluxes are obtained. A rigorous derivation provided in Evans and Morriss confirms this intuition~\cite{Evans_Morriss}. 
 
With these changes, the fluxes are related to the microscopic dynamics via
\begin{subequations}
\begin{align}
\label{eq:ik-j}
    \vq{j} &= \frac{1}{\mathcal{V}}\sum_{i=1}^N q_i \vq{v}_i,\\
\label{eq:ik-P}
    \vq{P} &= \frac{1}{\mathcal{V}}\sum_{i=1}^N\ \left( m_i \vq{v}_i \vq{v}_i + \frac{1}{2}\sum_{j \neq i}^N \vq{r}_{ij} \frac{\partial \phi_{ij}}{\partial \vq{r}_i} \right),\\
\label{eq:ik-q}
 \vq{q} &= \frac{1}{\mathcal{V}}\sum_{i=1}^N  \Biggr[ \vq{v}_i \left(\frac{1}{2}m_i \vq{v}_i^2 + \frac{1}{2}\sum_{j \neq i}^N \phi_{ij} \right)
    + \frac{1}{2} \sum_{j \neq i}^N (\vq{r}_i \cdot \vq{v}_i)  \frac{\partial \phi_{ij}}{\partial \vq{r}_i}\Biggr].
\end{align}
\end{subequations}
The pressure tensor is seen to contain contributions from the kinetic and interaction energy of a system's constituent particles. The heat flux contains three contributions, two of which stem from the kinetic and interaction energy that particles transport when they move, and another which represents a transfer of energy through collisions. 
Recall that it is the viscous pressure tensor $\hat{\vq{P}}$ that appears in the linear constitutive relation for viscosity. To obtain this, one can simply subtract the time-averaged value of the pressure tensor from its instantaneous value, i.e.,
\begin{equation}
    \hat{\vq{P}} = \vq{P} - \langle \vq{P} \rangle_t
\end{equation}
because thermodynamic fluxes are zero on average at equilibrium.

\subsection{Green-Kubo relations}
 Green-Kubo relations relate transport coefficients to fluctuations of thermodynamic fluxes in a system at equilibrium~\cite{Green_1954, kubo_1957}. They generally rely on Onsager's regression hypothesis~\cite{Onsager_1931}, which states that the average regression of equilibrium fluctuations obey the same laws as the corresponding macroscopic irreversible processes. They are commonly calculated from MD simulations in plasma physics,~\cite{HansenPRA1975,OttPRE2015,Scheiner_2020,DonkoPRL1998,DaligaultPRE2014} particularly in regimes where kinetic theory fails~\cite{transport_workshop}. Readers interested in a derivation should consult one of the many in literature~\cite{Green_1954, kubo_1957, Zwanzig_1964, Evans_Morriss, hansen_2013}. 

In terms of the phenomenological coefficients, the Green-Kubo relations are
\begin{equation}
    L_{ik} = \frac{\mathcal{V}}{k_B} \int_0^\infty dt \, \langle \mathcal{J}_i (t) \mathcal{J}_k(0) \rangle,
\end{equation}
where the angle brackets denote an ensemble average. 
Therefore, the second rank transport tensors of the previous section may be written
\begin{subequations}
\label{eq:green_kubo}
    \begin{eqnarray}
    \label{eq:sigma-gk}
        \bm{\sigma} = \frac{\mathcal{V}}{k_B T} &\int_0^\infty dt \, \langle \vq{j}(t) \vq{j}(0)\rangle \\
        \label{eq:varphi-gk}
        \bm{\varphi} = -\frac{\mathcal{V}}{k_B T^2} &\int_0^\infty dt \, \langle \vq{j}(t) \vq{q}(0)\rangle \\
        \label{eq:lambda-gk}
        \bm{\lambda} = \frac{\mathcal{V}}{k_B T^2} &\int_0^\infty dt \, \langle \vq{q}(t) \vq{q}(0)\rangle \\
        \label{eq:phi-gk}
        \bm{\phi} = \frac{\mathcal{V}}{k_B T} &\int_0^\infty dt \, \langle \vq{q}(t) \vq{j}(0)\rangle.
    \end{eqnarray}
\end{subequations}
The tensor of viscosity coefficients is written most concisely in its phenomenological form, i.e., 
\begin{equation}
\label{eq:gk-P}
   \vq{L}_{PP} = \frac{\mathcal{V}}{k_B} \int_0^\infty dt \langle \hat{\vq{P}}(t) \hat{\vq{P}}(0) \rangle.
\end{equation}
In practice, in order to compute from MD simulations, the autocorrelation must be discretized in time and cutoff after some long time $L$ by which the autocorrelation has decayed to zero. Additionally, the ensemble average is often replaced by a time average; one obtains a time-series of length $t_N \gg L$ and averages $t_N - L$ autocorrelations together. A general form of the Green-Kubo relations can therefore be written~\cite{Frenkel_book}
\begin{equation}
    L_{ik} = \frac{\mathcal{V}}{k_B} \frac{\Delta t}{t_N - L + 1} \sum_{\tau_t=0}^t \sum_{\tau = 0}^{t_N - L} \mathcal{J}_i(\tau + \tau_t) \mathcal{J}_k (\tau)
\end{equation}
where $\Delta t$ is the length of a timestep, $t_N$ is the total number of timesteps, and $L$ is the autocorrelation length. 

With this discussion, the fluid equations are nearly closed. Transport coefficients can be evaluated using the Green-Kubo relations, employing the fluxes derived in the Irving-Kirkwood procedure. Importantly, the results here are quite general and can be applied to a wide range of plasmas, including those which cannot yet be described using kinetic theory. 

\subsection{Equation of state\label{sec:eos}}
The last step towards obtaining single fluid equations is to determine expressions for the excess internal energy $u_{\mathrm{ex}}$, excess pressure $p_{\mathrm{ex}}$, and excess electron chemical potential $\mu_{e\mathrm{,ex}}$. There are a wide range of approaches to this problem in dense plasmas, but here, as with transport, we will consider a classical statistical physics approach. 

The radial distribution function between species $\alpha$ and $\beta$ is defined as~\cite{hansen_2013}
\begin{equation}
    g_{\alpha \beta}(r) = \frac{\langle n_{\alpha \beta}(r) \rangle}{n_\beta},
\end{equation}
which can be interpreted as the averaged radial density profile of $\beta$ particles around a reference particle of species $\alpha$, normalized by the background density. In terms of particle properties, this can be equivalently written
\begin{equation}
    g_{\alpha \beta}(r) = \frac{\mathcal{V}}{N_\alpha N_\beta} \sum_a^{N_\alpha} \sum_b^{N_\beta} \langle \delta(|\vq{r}_a - \vq{r}_b| - r) \rangle.
\end{equation}
Normalization ensures that for large distances, $g(r) \rightarrow 1$. 
The radial distribution function can be determined experimentally with scattering experiments~\cite{GlenzerRMP2009}, computationally with molecular dynamics~\cite{BrushJCP1966}, or theoretically in various approximations~\cite{hansen_2013}.
For plasmas, the hypernetted chain approximation is accurate in most conditions~\cite{Baus_1980}. 
In the weakly coupled limit, $g(r) = \exp(-\phi_{\textrm{DH}}/k_BT)$, where $\phi_{\textrm{DH}}$ is the Debye-H\"{u}ckel potential. 

Since $g(r)$ can often be modeled, it is useful to express the excess thermodynamic quantities in terms of it. For the excess internal energy and pressure, exact expressions can be derived from statistical physics~\cite{hansen_2013}
\begin{gather}
    u_{\mathrm{ex}} = 2 \pi \frac{n^2}{\rho} \sum_\alpha \sum_\beta x_\alpha x_\beta \int_0^\infty \phi_{\alpha \beta}(r) g_{\alpha \beta}(r) r^2 dr \\
    \label{eq:pex}
    p_{\mathrm{ex}} = -\frac{2}{3} \pi n^2 \sum_\alpha \sum_\beta x_\alpha x_\beta \int_0^\infty \frac{d\phi_{\alpha \beta}(r)}{dr} g_{\alpha \beta}(r) r^3 dr
\end{gather}
where $x_\alpha$ is the number fraction of species $\alpha$, $\phi_{\alpha \beta}(r)$ is the interaction potential, and the number density is $n = \sum_\alpha \rho_\alpha/m_\alpha$.

For the chemical potential, one cannot construct an exact expression in terms of $g(r)$. Physically, the radial distribution function only considers pair correlations, but the chemical potential contains contributions from $N$-particle correlations. 
There are a number of different approximations for the excess chemical potential in terms of $g(r)$ however, for example under the hypernetted chain approximation~\cite{Baus_1980}. Here we simply present a result that relates $\mu_{e,\mathrm{ex}}$ to $p_{\mathrm{ex}}$ through thermodynamic relations since an exact expression for $p_{\mathrm{ex}}$ has been presented. First, one can show from straightforward manipulation of the Gibbs function $G$ that~\cite{Tsednee_2019}
\begin{equation}
    \tilde{\mu}_{\mathrm{ex}} = \int_0^n dn' \frac{1}{n'} \left(\frac{\partial p_{\mathrm{ex}}}{\partial n'}\right)_T
\end{equation}
where $\tilde{\mu}_{\mathrm{ex}}$ is the excess thermodynamic (not specific) chemical potential. Then, from the definition of a partial specific thermodynamic quantity, one has $\mu_{e,\mathrm{ex}} = \partial \tilde{\mu}_{\mathrm{ex}} / \partial M_e$ where $M_e = m_e N_e$ so 
\begin{equation}
\label{eq:mue}
    \mu_{e,\mathrm{ex}} = \frac{\partial}{\partial M_e}  \int_0^n dn' \frac{1}{n'} \left(\frac{\partial p_{\mathrm{ex}}}{\partial n'}\right)_T,
\end{equation}
which can be evaluated using  $p_\mathrm{ex}$ from Eq.~(\ref{eq:pex}). 

There are several challenges associated with applying the approach outlined in this section to plasmas, however. 
First, using Eq.~(\ref{eq:pex}) for the excess pressure in Eq.~(\ref{eq:mue}) is only approximate because derivatives of $p_\mathrm{ex}$ with respect to thermodynamic variables depend on higher order correlation functions. 
More fundamentally, however, is that classical descriptions of dense electron-ion systems suffer thermodynamic instability, an effect that is sometimes called Coulomb collapse.~\cite{Baus_1980} 
This occurs because the attractive nature of the interaction allows particles to become arbitrarily close to one another, leading to thermodynamic instability and the divergence of excess components of the equation of state. 
Of course, this does not influence the ideal limit, which neglects these interactions entirely. 
It is well known that the stability of matter relies fundamentally on a quantum description at microscopic scales, specifically at scales at or below the thermal de Broglie wavelength.~\cite{Baus_1980,pathria2011} 
Semiclassical approximations can be made using the framework described in this section for moderately dense systems by modifying the electron-ion force at distances closer than the de Broglie wavelength to model quantum effects.~\cite{filinov2004temperature} 
This is called a pseudopotential technique. 
However, a fundamentally quantum mechanical approach is required to treat truly dense systems. 
Doing so in a first-principles manner is not yet possible, but is an active field of research.~\cite{DornheimPRB2025} 
See, for example, a recent review on dense hydrogen.~\cite{BonitzPOP2024}



\section{Discussion\label{sec:compare}}
Closed MHD equations have now been obtained. Before discussing common limits, we summarize the results. Expressions for conservation of mass, momentum, and energy were found to be 
\begin{gather*}
    \tag{\ref{eq:final-mass}}
    \frac{1}{\rho} \frac{d\rho}{dt} = - \nabla \cdot \vq{V} \\
    \tag{\ref{eq:final-mom}}
    \rho \frac{d\vq{V}}{dt} = - \nabla \cdot \vq{P} + \vq{j} \times \vq{B} \\ 
    \tag{\ref{eq:final-energy}}
    \rho \frac{du}{dt} = - \nabla \cdot \vq{q} - \vq{P}:\nabla \vq{V} + \vq{j} \cdot \vq{E}'.
\end{gather*}
These are coupled with Maxwell's equations, which in a non-relativistic quasi-neutral plasma are
\begin{subequations}
\begin{align}
    \tag{\ref{eq:maxa}}
    \nabla \times \vq{E} &= - \frac{\partial \vq{B}}{\partial t} \\ 
    \tag{\ref{eq:maxb}}
    \nabla \times \vq{B} &= \mu_0 \vq{j}, \\
    \tag{\ref{eq:maxc}}
    \nabla \cdot \vq{B} &= 0 .
\end{align}
\end{subequations}
After analyzing the entropy production of an ion-electron plasma, the following linear constitutive relations for current density, heat flux, and viscous pressure were established:
\begin{gather*}
    \tag{\ref{eq:current-final}}
    \vq{j} = \bm{\sigma} \cdot \left[\vq{E}' - \frac{T}{z_e} \nabla \frac{\mu_e}{T} \right] + \bm{\varphi} \cdot \nabla T, \\
    \tag{\ref{eq:heat-final}}
    \vq{q} = -\bm{\lambda} \cdot \nabla T + \bm{\phi} \cdot \left[\vq{E}' - \frac{T}{z_e} \nabla \frac{\mu_e}{T} \right], \\
    \tag{\ref{eq:final-visc}}
    \hat{\vq{P}} = - \vq{L}_{PP} : \frac{\nabla \vq{V}}{T}.
\end{gather*}
Symmetry arguments were used to show that the rank two tensors of transport coefficients can be written as
\begin{equation}
\tag{\ref{eq:gen-rank2}}
    \vq{L} = 
    \begin{pmatrix}
        L_{\perp} & -L_{\wedge} & 0\\
        L_{\wedge} & L_{\perp} & 0\\
        0 & 0 & L_{\parallel}
    \end{pmatrix},
\end{equation}
i.e., they are each composed of just three independent coefficients, and that $\bm{\phi}(\vq{B}) = - \bm{\varphi}^\top (-\vq{B}) T$. Additionally, it was shown that there are seven independent viscosity coefficients and that they can be organized in a similar way with orientation to the magnetic field as described in Eq.~(\ref{eq:P_final}). To fully close the MHD equations, methods were developed to compute the transport coefficients and equations of state ($u$, $p$, and $\mu_e$) from particle trajectories, as described by the Green-Kubo relations in Eq.~(\ref{eq:green_kubo}) and Sec.~\ref{sec:eos}.  

This constituents a general single fluid description. 
Many different versions of MHD equations are quoted in literature, but all should correspond to approximate forms of these general equations. 
We summarize a few examples of common limiting cases next. 

\subsection{Weakly magnetized limit}
In the limit of a weak magnetic field, symmetry arguments can be used to reduce the transport tensors to scalars. Recall the orthogonal transformations discussed in Sec.~\ref{sec-trans}. With a weak magnetic field, rotational symmetry is guaranteed about any axis, as opposed to just the magnetic field axis. Hence, one can apply a small rotation about the x-axis
\begin{equation}
    \vq{A}_3 = 
    \begin{pmatrix}
        1 & 0 & 0\\
        0 & 1 & -\alpha\\
        0 & \alpha & 1
    \end{pmatrix}
\end{equation}
to the transport tensors and conclude that, for a rank two tensor, $L_\parallel$ = $L_\perp$ and $L_\wedge = 0$. Similar arguments can be applied to the rank four viscosity tensor, and it is found that just two coefficients, one corresponding to shear viscosity and the other to bulk viscosity, are independent~\cite{degroot_mazur}. The following scalar representation is therefore obtained
\begin{subequations}
\begin{gather}
    \vq{j} = \sigma\left[\vq{E}' - \frac{T}{z_e} \nabla \frac{\mu_e}{T} \right] + \varphi \nabla T, \\
    \vq{q} = - \lambda \nabla T +\phi\left[\vq{E}' - \frac{T}{z_e} \nabla \frac{\mu_e}{T} \right], \\
    \hat{\vq{P}} = - \eta \left[\nabla \vq{V} + (\nabla \vq{V})^{\top} - \frac{2}{3}(\nabla \cdot \vq{V})\vq{I}\right] - \eta_v (\nabla \cdot \vq{V}) \vq{I}.
\end{gather}
\end{subequations}
The equation of state is unaffected, owing to the Bohr-Van Leeuwen theorem~\cite{van_leeuwen}.

\subsection{Ideal gas limit: Chapman-Enskog result}
The ideal gas limit of the equations obtained here matches the Chapman-Enskog solution. Recall that in the Chapman-Enskog method, mass, momentum and energy moments of the Boltzmann equation are taken to arrive at MHD equations~\cite{chapman_cowling}. Transport coefficients can be calculated directly from the plasma density and temperature, but the results are only valid for dilute plasmas. Here, we will compare with the solution obtained by Ferziger and Kaper~\cite{ferziger_kaper}. It is worth noting that even among the standard references on this procedure,~\cite{chapman_cowling, ferziger_kaper, braginskii_review, balescu_text} there are differences in the transport equations obtained by the various authors. See the text by Balescu for details.~\cite{balescu_text} 

First, the form of the conservation laws is the same as Eqs.~(\ref{eq:final-mass}), (\ref{eq:final-mom}) and (\ref{eq:final-energy}), so long as the pressure and internal energy are related to density and temperature via the ideal gas equation of state: $p = nk_B T$ and $\rho u = \frac{3}{2} nk_B T$. 
However, the linear constitutive relations for heat flux and current density differ by a thermodynamic factor. In particular, the Chapman-Enskog result is~\cite{chapman_cowling, ferziger_kaper}
\begin{subequations}
    \begin{align}
    \label{eq:j_ce}
        \vq{j} &= \bm{\sigma} \cdot \left(\vq{E}' + \frac{1}{en} \nabla p  \right) + \bm{\varphi}^\prime \cdot \nabla T \\
        \label{eq:q_ce}
        \vq{q} &= - \bm{\lambda}' \cdot \nabla T + \bm{\psi}_C \cdot \vq{j},
    \end{align}
\end{subequations}
where $\bm{\varphi}^\prime$ is an electrothermal coefficient with a different definition from above, $\bm{\lambda}'$ is the ``total'' thermal conductivity and $\bm{\psi}_C$ is a coefficient related to thermal diffusion. For comparison, the general results presented here are Eqs.~(\ref{eq:current-final}) and (\ref{eq:heat-final}).
It is straightforward to show that, in the ideal gas limit, the expressions are equivalent. 
First, from Eq.~(\ref{eq:mu_e_ideal}) for an ideal gas, 
\begin{equation}
     \nabla \frac{\mu_e}{T} = \frac{k_B}{m_e p} \nabla p - \frac{5}{2} \frac{k_B}{m_e T} \nabla T .
\end{equation}
Insertion of this into the general current density expression from Eq.~(\ref{eq:current-final}) yields the Champman-Enskog version of Ohm's law from Eq.~(\ref{eq:j_ce}) if
\begin{equation}
\bm{\varphi}^\prime  = \bm{\varphi} - \frac{5k_B}{2e} \bm{\sigma} .    
\end{equation}
Similarly, the  general heat flux expression from Eq.~(\ref{eq:heat-final}) agrees with the Chapman-Enskog expression from Eq.~(\ref{eq:q_ce}) if 
\begin{equation}
    \bm{\lambda}'=\bm{\lambda} + \bm{\phi} \cdot \bm{\sigma}^{-1} \cdot \bm{\varphi}   \quad \mathrm{and} \quad \bm{\psi}_C = \bm{\phi} \cdot \bm{\sigma}^{-1}.
\end{equation} 
The linear constitutive relation for viscosity was already written in such a way to match the Chapman-Enskog result as reviewed by Braginskii~\cite{braginskii_review}. If one sets the cross viscosity ($\zeta$) and bulk viscosity ($\eta_v$) terms to zero, which is true in the monatomic ideal gas limit, then Eq.~(\ref{eq:newton}) clearly matches the Chapman-Enskog result. 

\subsection{Generalized Ohm's Law}

There are many ``extended'' MHD models in the literature. These attempt to capture so-called two-fluid effects through a generalized Ohm's law. 
For example, it is common to see the following form~\cite{gurnett_2005}
\begin{align}
\label{eq:gen-ohm}
    \vq{E} + \vq{v} \times \vq{B} = \eta \vq{j} + \frac{1}{en_e} \vq{j} \times \vq{B} + \frac{1}{en_e} \nabla \cdot \vq{P}_e \\ \nonumber + \frac{m_e}{en_e} \left[\frac{\partial \vq{j}}{\partial t} + \nabla \cdot (\vq{j}\vq{v} + \vq{v}\vq{j}) \right],
\end{align}
where $\eta = 1/\sigma$ is a scalar coefficient of resistivity. This is derived from two-fluid ion and electron momentum equations. 
Here, it is argued that terms often referred to as two-fluid effects are also present in the single fluid model, and that this is not necessarily more general than the single fluid description described above. 
To compare the two equations, define the resistivity tensor $\bm{\eta} = \bm{\sigma}^{-1}$ and expand $\vq{E}' = \vq{E} + \vq{v} \times \vq{B}$ to write Eq.~(\ref{eq:current-final}) as
\begin{equation}
\label{eq:ohm_1}
    \vq{E} + \vq{v} \times \vq{B} = \bm{\eta} \cdot \vq{j}  + \frac{T}{z_e} \nabla \frac{\mu_e}{T} - \bm{\eta} \cdot \bm{\varphi} \cdot \nabla T.
\end{equation}
Several differences between Eqs.~(\ref{eq:gen-ohm}) and (\ref{eq:ohm_1}) should be noted. 

First, the extended Ohm's law of Eq.~(\ref{eq:gen-ohm}) is not closed because the pressure tensor has not been related to the state variables. 
If the pressure tensor is modeled only as a scalar pressure, as is often done, then the expression misses all electrothermal effects. 
The chemical potential and temperature gradient terms in Eq.~(\ref{eq:ohm_1}) should somehow map to this, at least in the weakly coupled limit, but without a closure, it is impossible to compare the two. 

Second, the resistivity terms are different. 
Resistivity is a tensor with three independent coefficients in Eq.~(\ref{eq:ohm_1}), $\bm{\eta} \cdot \bm{j} = \eta_\parallel \bm{j}_\parallel +\eta_\perp \bm{j}_\perp + \eta_\wedge \bm{j}_\wedge$, but has only a single scalar coefficient in Eq.~(\ref{eq:gen-ohm}) plus the ``Hall'' $\bm{j}\times\bm{B}$ term.  
The first two terms of the general expression return $\eta \bm{j}$ if $\eta_\parallel = \eta_\perp$.
However, even in the Chapman-Enskog solution it is known that $\eta_\perp \approx 1.9 \eta_\parallel$ if the plasma is magnetized,~\cite{ferziger_kaper} so this approximation loses information about asymmetry in the resistivity. 
Regarding the Hall term, which is often called a two-fluid effect, we note that this is captured in the off-diagonal ($\eta_\wedge$) element of the resistivity tensor. 
In a Chapman-Enskog solution, one finds $\eta_\wedge \bm{j}_\wedge \approx \bm{j}\times \bm{B}/(en_e)$. 
So rather than being a two-fluid effect, Hall physics is also captured in the single-fluid MHD. 

Third, the generalized Ohm's law in Eq.~(\ref{eq:gen-ohm}) includes an inertial term [last line of Eq.~(\ref{eq:gen-ohm})]. This is a two-fluid effect that does not appear in the Ohm's law developed in this tutorial. However, it can be recovered within the same framework. 
Recall from Eq.~(\ref{eq:e-density}) that the kinetic energy of diffusion $\sum_\alpha \vq{d}_\alpha^2 / 2 \rho_\alpha$ was ignored. If one instead kept this term and propagated it through the non-equilibrium thermodynamics arguments, using the fact that $\sum_\alpha \vq{d}_\alpha^2 / 2 \rho_\alpha \approx m_e \vq{j}^2 / (2e^2 n_e)$, the inertial term can be recovered. However, for a one fluid model, the diffusive flux of any species must be small. If the diffusive kinetic energy of a species is significant, single fluid hydrodynamics is not applicable. The inertial term can be considered a two-fluid effect that is second-order in the linear expansion from equilibrium.  

\subsection{Resistive and ideal MHD}

In plasma physics, ``MHD'' is often used synonymously with resistive MHD.  
Resistive MHD can be obtained from a limit of the general equations presented here if all transport coefficients are set to zero except a single scalar resistivity coefficient in Ohm's law. 
All viscosity, thermal conduction, and Ohmic heating effects are thus assumed negligible, setting the following terms to zero: $\hat{\vq{P}}$, $\vq{q}$, $\nabla T$, and $\vq{j} \cdot \vq{E}'$. Additionally, one assumes an ideal gas equation of state and with regards to transport, assumes a weakly magnetized limit, including the neglect of Hall physics.  

With these changes, the only dissipative process considered is the flow of current. The conservation equations in resistive MHD equations become~\cite{gurnett_2005}
\begin{subequations}
    \begin{eqnarray}
        \frac{1}{\rho} \frac{d\rho}{dt} = - \nabla \cdot \vq{V} \\
        \rho \frac{d\vq{V}}{dt} = - \nabla p + \vq{j} \times \vq{B} \\
        \frac{d}{dt}\left( \frac{p}{\rho^\gamma}\right) = 0 
    \end{eqnarray}
\end{subequations}
where $\gamma$ is the polytropic index. The mass conservation equation is seen to be unchanged, but momentum and energy equations are far simpler. Ohm's law in resistive MHD is simply
\begin{equation}
    \vq{E} + \vq{V} \times \vq{B} = \eta \vq{j},
\end{equation}
which, coupled with the standard pre-Maxwell equations, forms a closed set of equations. 

A further simplification is made in ideal MHD. One keeps the assumptions of resistive MHD, and also assumes resistive effects are negligible ($\eta = 0$). Ohm's law in ideal MHD is thus
\begin{equation}
    \vq{E} + \vq{V} \times \vq{B} = 0.
\end{equation}
For a discussion on when the equations of resistive and ideal MHD are expected to be valid, one should consult introductory texts on plasma physics\cite{gurnett_2005,Boyd_Sanderson_2003,Freidberg_book}. 

\section{Conclusions}
This tutorial outlined a derivation of single fluid MHD equations valid beyond the ideal gas limit. Conservation laws and linear constitutive relations for transport were obtained with non-equilibrium thermodynamics. Symmetry arguments were then used to vastly reduce the number of independent coefficients appearing in the transport equations. This procedure yields the structure of the MHD equations, but lacks a means to compute the transport coefficients and equations of state. For these values, a connection to microscopic dynamics was developed, enabling one to compute them from particle trajectories in an equilibrium system.

Assumptions were generally kept to a minimum, so the obtained equations are quite general. 
It is expected to apply to dilute or dense plasmas, as well as liquid metals. 
This makes it useful to compare the equations here to common forms of MHD obtained from plasma kinetic theories, in order to get an understanding of when the latter are appropriate. Additionally, the equations developed are valid for dense systems, which do not satisfy an ideal gas equation of state. This leads to an electron chemical potential term in Ohm's and Fourier's law, motivating a need for accurate models of this quantity in dense plasmas.


%

\begin{acknowledgments}
The authors thank Lucas Babati for helpful conversations on this work. 
This research was supported by the US National Science Foundation under Award No.~PHY-2205506, and by the NNSA Stewardship Science Academic Programs under DOE Cooperative Agreement DE-NA0004148. 
\end{acknowledgments}

\section*{Author declarations}
The authors have no conflicts to disclose.

\section*{Data Availability Statement}
Data sharing is not applicable to this article as no new data were created or analyzed in this study. 

\bibliography{aipsamp}

\providecommand{\noopsort}[1]{}\providecommand{\singleletter}[1]{#1}%
\begin{thebibliography}{50}%
\makeatletter
\providecommand \@ifxundefined [1]{%
 \@ifx{#1\undefined}
}%
\providecommand \@ifnum [1]{%
 \ifnum #1\expandafter \@firstoftwo
 \else \expandafter \@secondoftwo
 \fi
}%
\providecommand \@ifx [1]{%
 \ifx #1\expandafter \@firstoftwo
 \else \expandafter \@secondoftwo
 \fi
}%
\providecommand \natexlab [1]{#1}%
\providecommand \enquote  [1]{``#1''}%
\providecommand \bibnamefont  [1]{#1}%
\providecommand \bibfnamefont [1]{#1}%
\providecommand \citenamefont [1]{#1}%
\providecommand \href@noop [0]{\@secondoftwo}%
\providecommand \href [0]{\begingroup \@sanitize@url \@href}%
\providecommand \@href[1]{\@@startlink{#1}\@@href}%
\providecommand \@@href[1]{\endgroup#1\@@endlink}%
\providecommand \@sanitize@url [0]{\catcode `\\12\catcode `\$12\catcode `\&12\catcode `\#12\catcode `\^12\catcode `\_12\catcode `\%12\relax}%
\providecommand \@@startlink[1]{}%
\providecommand \@@endlink[0]{}%
\providecommand \url  [0]{\begingroup\@sanitize@url \@url }%
\providecommand \@url [1]{\endgroup\@href {#1}{\urlprefix }}%
\providecommand \urlprefix  [0]{URL }%
\providecommand \Eprint [0]{\href }%
\providecommand \doibase [0]{http://dx.doi.org/}%
\providecommand \selectlanguage [0]{\@gobble}%
\providecommand \bibinfo  [0]{\@secondoftwo}%
\providecommand \bibfield  [0]{\@secondoftwo}%
\providecommand \translation [1]{[#1]}%
\providecommand \BibitemOpen [0]{}%
\providecommand \bibitemStop [0]{}%
\providecommand \bibitemNoStop [0]{.\EOS\space}%
\providecommand \EOS [0]{\spacefactor3000\relax}%
\providecommand \BibitemShut  [1]{\csname bibitem#1\endcsname}%
\let\auto@bib@innerbib\@empty
\bibitem [{\citenamefont {Freidberg}(1982)}]{Freidberg_1982}%
  \BibitemOpen
  \bibfield  {author} {\bibinfo {author} {\bibfnamefont {J.~P.}\ \bibnamefont {Freidberg}},\ }\href {\doibase 10.1103/RevModPhys.54.801} {\bibfield  {journal} {\bibinfo  {journal} {Rev. Mod. Phys.}\ }\textbf {\bibinfo {volume} {54}},\ \bibinfo {pages} {801} (\bibinfo {year} {1982})}\BibitemShut {NoStop}%
\bibitem [{\citenamefont {Alfv{\'e}n}(1942)}]{Alfven_1942}%
  \BibitemOpen
  \bibfield  {author} {\bibinfo {author} {\bibfnamefont {H.}~\bibnamefont {Alfv{\'e}n}},\ }\href@noop {} {\bibfield  {journal} {\bibinfo  {journal} {Nature}\ }\textbf {\bibinfo {volume} {150}},\ \bibinfo {pages} {405} (\bibinfo {year} {1942})}\BibitemShut {NoStop}%
\bibitem [{\citenamefont {Pontin}\ and\ \citenamefont {Priest}(2022)}]{recon_rev}%
  \BibitemOpen
  \bibfield  {author} {\bibinfo {author} {\bibfnamefont {D.~I.}\ \bibnamefont {Pontin}}\ and\ \bibinfo {author} {\bibfnamefont {E.~R.}\ \bibnamefont {Priest}},\ }\href@noop {} {\bibfield  {journal} {\bibinfo  {journal} {Living Reviews in Solar Physics}\ }\textbf {\bibinfo {volume} {19}},\ \bibinfo {pages} {1} (\bibinfo {year} {2022})}\BibitemShut {NoStop}%
\bibitem [{\citenamefont {Beresnyak}(2019)}]{beresnyak_2019}%
  \BibitemOpen
  \bibfield  {author} {\bibinfo {author} {\bibfnamefont {A.}~\bibnamefont {Beresnyak}},\ }\href@noop {} {\bibfield  {journal} {\bibinfo  {journal} {Living Reviews in Computational Astrophysics}\ }\textbf {\bibinfo {volume} {5}},\ \bibinfo {pages} {2} (\bibinfo {year} {2019})}\BibitemShut {NoStop}%
\bibitem [{\citenamefont {Hawley}(2003)}]{Hawley_2003}%
  \BibitemOpen
  \bibfield  {author} {\bibinfo {author} {\bibfnamefont {J.~F.}\ \bibnamefont {Hawley}},\ }\href {\doibase 10.1063/1.1542885} {\bibfield  {journal} {\bibinfo  {journal} {Physics of Plasmas}\ }\textbf {\bibinfo {volume} {10}},\ \bibinfo {pages} {1946} (\bibinfo {year} {2003})}\BibitemShut {NoStop}%
\bibitem [{\citenamefont {Burlaga}(1984)}]{burlaga1984mhd}%
  \BibitemOpen
  \bibfield  {author} {\bibinfo {author} {\bibfnamefont {L.}~\bibnamefont {Burlaga}},\ }\href@noop {} {\bibfield  {journal} {\bibinfo  {journal} {Space science reviews}\ }\textbf {\bibinfo {volume} {39}},\ \bibinfo {pages} {255} (\bibinfo {year} {1984})}\BibitemShut {NoStop}%
\bibitem [{\citenamefont {Chapman}\ and\ \citenamefont {Cowling}(1990)}]{chapman_cowling}%
  \BibitemOpen
  \bibfield  {author} {\bibinfo {author} {\bibfnamefont {S.}~\bibnamefont {Chapman}}\ and\ \bibinfo {author} {\bibfnamefont {T.~G.}\ \bibnamefont {Cowling}},\ }\href@noop {} {\emph {\bibinfo {title} {The mathematical theory of non-uniform gases: an account of the kinetic theory of viscosity, thermal conduction and diffusion in gases}}}\ (\bibinfo  {publisher} {Cambridge university press},\ \bibinfo {year} {1990})\BibitemShut {NoStop}%
\bibitem [{\citenamefont {Braginskii}(1965)}]{braginskii_review}%
  \BibitemOpen
  \bibfield  {author} {\bibinfo {author} {\bibfnamefont {S.}~\bibnamefont {Braginskii}},\ }\href@noop {} {\bibfield  {journal} {\bibinfo  {journal} {Reviews of plasma physics}\ }\textbf {\bibinfo {volume} {1}},\ \bibinfo {pages} {205} (\bibinfo {year} {1965})}\BibitemShut {NoStop}%
\bibitem [{\citenamefont {De~Groot}\ and\ \citenamefont {Mazur}(2013)}]{degroot_mazur}%
  \BibitemOpen
  \bibfield  {author} {\bibinfo {author} {\bibfnamefont {S.~R.}\ \bibnamefont {De~Groot}}\ and\ \bibinfo {author} {\bibfnamefont {P.}~\bibnamefont {Mazur}},\ }\href@noop {} {\emph {\bibinfo {title} {Non-equilibrium thermodynamics}}}\ (\bibinfo  {publisher} {Courier Corporation},\ \bibinfo {year} {2013})\BibitemShut {NoStop}%
\bibitem [{\citenamefont {Onsager}(1931)}]{Onsager_1931}%
  \BibitemOpen
  \bibfield  {author} {\bibinfo {author} {\bibfnamefont {L.}~\bibnamefont {Onsager}},\ }\href {\doibase 10.1103/PhysRev.37.405} {\bibfield  {journal} {\bibinfo  {journal} {Phys. Rev.}\ }\textbf {\bibinfo {volume} {37}},\ \bibinfo {pages} {405} (\bibinfo {year} {1931})}\BibitemShut {NoStop}%
\bibitem [{\citenamefont {Casimir}(1945)}]{Casimir_1945}%
  \BibitemOpen
  \bibfield  {author} {\bibinfo {author} {\bibfnamefont {H.~B.~G.}\ \bibnamefont {Casimir}},\ }\href {\doibase 10.1103/RevModPhys.17.343} {\bibfield  {journal} {\bibinfo  {journal} {Rev. Mod. Phys.}\ }\textbf {\bibinfo {volume} {17}},\ \bibinfo {pages} {343} (\bibinfo {year} {1945})}\BibitemShut {NoStop}%
\bibitem [{\citenamefont {Irving}\ and\ \citenamefont {Kirkwood}(1950)}]{Irving_Kirkwood}%
  \BibitemOpen
  \bibfield  {author} {\bibinfo {author} {\bibfnamefont {J.~H.}\ \bibnamefont {Irving}}\ and\ \bibinfo {author} {\bibfnamefont {J.~G.}\ \bibnamefont {Kirkwood}},\ }\href {\doibase 10.1063/1.1747782} {\bibfield  {journal} {\bibinfo  {journal} {The Journal of Chemical Physics}\ }\textbf {\bibinfo {volume} {18}},\ \bibinfo {pages} {817} (\bibinfo {year} {1950})}\BibitemShut {NoStop}%
\bibitem [{\citenamefont {Green}(1954)}]{Green_1954}%
  \BibitemOpen
  \bibfield  {author} {\bibinfo {author} {\bibfnamefont {M.~S.}\ \bibnamefont {Green}},\ }\href {\doibase 10.1063/1.1740082} {\bibfield  {journal} {\bibinfo  {journal} {The Journal of Chemical Physics}\ }\textbf {\bibinfo {volume} {22}},\ \bibinfo {pages} {398} (\bibinfo {year} {1954})}\BibitemShut {NoStop}%
\bibitem [{\citenamefont {Kubo}, \citenamefont {Yokota},\ and\ \citenamefont {Nakajima}(1957)}]{kubo_1957}%
  \BibitemOpen
  \bibfield  {author} {\bibinfo {author} {\bibfnamefont {R.}~\bibnamefont {Kubo}}, \bibinfo {author} {\bibfnamefont {M.}~\bibnamefont {Yokota}}, \ and\ \bibinfo {author} {\bibfnamefont {S.}~\bibnamefont {Nakajima}},\ }\href@noop {} {\bibfield  {journal} {\bibinfo  {journal} {Journal of the Physical Society of Japan}\ }\textbf {\bibinfo {volume} {12}},\ \bibinfo {pages} {1203} (\bibinfo {year} {1957})}\BibitemShut {NoStop}%
\bibitem [{\citenamefont {Stanek}\ \emph {et~al.}(2024)\citenamefont {Stanek}, \citenamefont {Kononov}, \citenamefont {Hansen}, \citenamefont {Haines}, \citenamefont {Hu}, \citenamefont {Knapp}, \citenamefont {Murillo}, \citenamefont {Stanton}, \citenamefont {Whitley}, \citenamefont {Baalrud}, \citenamefont {Babati}, \citenamefont {Baczewski}, \citenamefont {Bethkenhagen}, \citenamefont {Blanchet}, \citenamefont {Clay}, \citenamefont {Cochrane}, \citenamefont {Collins}, \citenamefont {Dumi}, \citenamefont {Faussurier}, \citenamefont {French}, \citenamefont {Johnson}, \citenamefont {Karasiev}, \citenamefont {Kumar}, \citenamefont {Lentz}, \citenamefont {Melton}, \citenamefont {Nichols}, \citenamefont {Petrov}, \citenamefont {Recoules}, \citenamefont {Redmer}, \citenamefont {Röpke}, \citenamefont {Schörner}, \citenamefont {Shaffer}, \citenamefont {Sharma}, \citenamefont {Silvestri}, \citenamefont {Soubiran}, \citenamefont {Suryanarayana}, \citenamefont {Tacu}, \citenamefont {Townsend},\ and\ \citenamefont
  {White}}]{transport_workshop}%
  \BibitemOpen
  \bibfield  {author} {\bibinfo {author} {\bibfnamefont {L.~J.}\ \bibnamefont {Stanek}}, \bibinfo {author} {\bibfnamefont {A.}~\bibnamefont {Kononov}}, \bibinfo {author} {\bibfnamefont {S.~B.}\ \bibnamefont {Hansen}}, \bibinfo {author} {\bibfnamefont {B.~M.}\ \bibnamefont {Haines}}, \bibinfo {author} {\bibfnamefont {S.~X.}\ \bibnamefont {Hu}}, \bibinfo {author} {\bibfnamefont {P.~F.}\ \bibnamefont {Knapp}}, \bibinfo {author} {\bibfnamefont {M.~S.}\ \bibnamefont {Murillo}}, \bibinfo {author} {\bibfnamefont {L.~G.}\ \bibnamefont {Stanton}}, \bibinfo {author} {\bibfnamefont {H.~D.}\ \bibnamefont {Whitley}}, \bibinfo {author} {\bibfnamefont {S.~D.}\ \bibnamefont {Baalrud}}, \bibinfo {author} {\bibfnamefont {L.~J.}\ \bibnamefont {Babati}}, \bibinfo {author} {\bibfnamefont {A.~D.}\ \bibnamefont {Baczewski}}, \bibinfo {author} {\bibfnamefont {M.}~\bibnamefont {Bethkenhagen}}, \bibinfo {author} {\bibfnamefont {A.}~\bibnamefont {Blanchet}}, \bibinfo {author} {\bibfnamefont {I.}~\bibnamefont {Clay}, \bibfnamefont
  {Raymond~C.}}, \bibinfo {author} {\bibfnamefont {K.~R.}\ \bibnamefont {Cochrane}}, \bibinfo {author} {\bibfnamefont {L.~A.}\ \bibnamefont {Collins}}, \bibinfo {author} {\bibfnamefont {A.}~\bibnamefont {Dumi}}, \bibinfo {author} {\bibfnamefont {G.}~\bibnamefont {Faussurier}}, \bibinfo {author} {\bibfnamefont {M.}~\bibnamefont {French}}, \bibinfo {author} {\bibfnamefont {Z.~A.}\ \bibnamefont {Johnson}}, \bibinfo {author} {\bibfnamefont {V.~V.}\ \bibnamefont {Karasiev}}, \bibinfo {author} {\bibfnamefont {S.}~\bibnamefont {Kumar}}, \bibinfo {author} {\bibfnamefont {M.~K.}\ \bibnamefont {Lentz}}, \bibinfo {author} {\bibfnamefont {C.~A.}\ \bibnamefont {Melton}}, \bibinfo {author} {\bibfnamefont {K.~A.}\ \bibnamefont {Nichols}}, \bibinfo {author} {\bibfnamefont {G.~M.}\ \bibnamefont {Petrov}}, \bibinfo {author} {\bibfnamefont {V.}~\bibnamefont {Recoules}}, \bibinfo {author} {\bibfnamefont {R.}~\bibnamefont {Redmer}}, \bibinfo {author} {\bibfnamefont {G.}~\bibnamefont {Röpke}}, \bibinfo {author} {\bibfnamefont
  {M.}~\bibnamefont {Schörner}}, \bibinfo {author} {\bibfnamefont {N.~R.}\ \bibnamefont {Shaffer}}, \bibinfo {author} {\bibfnamefont {V.}~\bibnamefont {Sharma}}, \bibinfo {author} {\bibfnamefont {L.~G.}\ \bibnamefont {Silvestri}}, \bibinfo {author} {\bibfnamefont {F.}~\bibnamefont {Soubiran}}, \bibinfo {author} {\bibfnamefont {P.}~\bibnamefont {Suryanarayana}}, \bibinfo {author} {\bibfnamefont {M.}~\bibnamefont {Tacu}}, \bibinfo {author} {\bibfnamefont {J.~P.}\ \bibnamefont {Townsend}}, \ and\ \bibinfo {author} {\bibfnamefont {A.~J.}\ \bibnamefont {White}},\ }\href {\doibase 10.1063/5.0198155} {\bibfield  {journal} {\bibinfo  {journal} {Physics of Plasmas}\ }\textbf {\bibinfo {volume} {31}},\ \bibinfo {pages} {052104} (\bibinfo {year} {2024})}\BibitemShut {NoStop}%
\bibitem [{\citenamefont {LeVan}, \citenamefont {Acciarri},\ and\ \citenamefont {Baalrud}(2024)}]{LeVan_2024}%
  \BibitemOpen
  \bibfield  {author} {\bibinfo {author} {\bibfnamefont {J.}~\bibnamefont {LeVan}}, \bibinfo {author} {\bibfnamefont {M.~D.}\ \bibnamefont {Acciarri}}, \ and\ \bibinfo {author} {\bibfnamefont {S.~D.}\ \bibnamefont {Baalrud}},\ }\href {\doibase 10.1103/PhysRevE.110.015208} {\bibfield  {journal} {\bibinfo  {journal} {Phys. Rev. E}\ }\textbf {\bibinfo {volume} {110}},\ \bibinfo {pages} {015208} (\bibinfo {year} {2024})}\BibitemShut {NoStop}%
\bibitem [{\citenamefont {LeVan}\ and\ \citenamefont {Baalrud}(2025)}]{LeVan_2025}%
  \BibitemOpen
  \bibfield  {author} {\bibinfo {author} {\bibfnamefont {J.}~\bibnamefont {LeVan}}\ and\ \bibinfo {author} {\bibfnamefont {S.~D.}\ \bibnamefont {Baalrud}},\ }\href {\doibase 10.1103/PhysRevE.111.015202} {\bibfield  {journal} {\bibinfo  {journal} {Phys. Rev. E}\ }\textbf {\bibinfo {volume} {111}},\ \bibinfo {pages} {015202} (\bibinfo {year} {2025})}\BibitemShut {NoStop}%
\bibitem [{\citenamefont {Bearman}\ and\ \citenamefont {Kirkwood}(1958)}]{bearman_1958}%
  \BibitemOpen
  \bibfield  {author} {\bibinfo {author} {\bibfnamefont {R.~J.}\ \bibnamefont {Bearman}}\ and\ \bibinfo {author} {\bibfnamefont {J.~G.}\ \bibnamefont {Kirkwood}},\ }\href {\doibase 10.1063/1.1744056} {\bibfield  {journal} {\bibinfo  {journal} {The Journal of Chemical Physics}\ }\textbf {\bibinfo {volume} {28}},\ \bibinfo {pages} {136} (\bibinfo {year} {1958})}\BibitemShut {NoStop}%
\bibitem [{\citenamefont {Maugin}(1993)}]{Maugin_1993}%
  \BibitemOpen
  \bibfield  {author} {\bibinfo {author} {\bibfnamefont {G.~A.}\ \bibnamefont {Maugin}},\ }\enquote {\bibinfo {title} {Non-equilibrium thermodynamics of electromagnetic solids},}\ in\ \href {\doibase 10.1007/978-3-7091-4321-6_4} {\emph {\bibinfo {booktitle} {Non-Equilibrium Thermodynamics with Application to Solids: Dedicated to the Memory of Professor Theodor Lehmann}}},\ \bibinfo {editor} {edited by\ \bibinfo {editor} {\bibfnamefont {W.}~\bibnamefont {Muschik}}}\ (\bibinfo  {publisher} {Springer Vienna},\ \bibinfo {address} {Vienna},\ \bibinfo {year} {1993})\ pp.\ \bibinfo {pages} {205--258}\BibitemShut {NoStop}%
\bibitem [{\citenamefont {Wolff}\ and\ \citenamefont {Albano}(1979)}]{Wolff_1979}%
  \BibitemOpen
  \bibfield  {author} {\bibinfo {author} {\bibfnamefont {P.}~\bibnamefont {Wolff}}\ and\ \bibinfo {author} {\bibfnamefont {A.}~\bibnamefont {Albano}},\ }\href {\doibase https://doi.org/10.1016/0378-4371(79)90149-3} {\bibfield  {journal} {\bibinfo  {journal} {Physica A: Statistical Mechanics and its Applications}\ }\textbf {\bibinfo {volume} {98}},\ \bibinfo {pages} {491} (\bibinfo {year} {1979})}\BibitemShut {NoStop}%
\bibitem [{\citenamefont {Kluitenberg}(1973)}]{Kluitenberg_1973}%
  \BibitemOpen
  \bibfield  {author} {\bibinfo {author} {\bibfnamefont {G.}~\bibnamefont {Kluitenberg}},\ }\href {\doibase https://doi.org/10.1016/0031-8914(73)90131-6} {\bibfield  {journal} {\bibinfo  {journal} {Physica}\ }\textbf {\bibinfo {volume} {68}},\ \bibinfo {pages} {75} (\bibinfo {year} {1973})}\BibitemShut {NoStop}%
\bibitem [{\citenamefont {Curie}(1908)}]{curie_1908}%
  \BibitemOpen
  \bibfield  {author} {\bibinfo {author} {\bibfnamefont {P.}~\bibnamefont {Curie}},\ }\href@noop {} {\emph {\bibinfo {title} {Works of Pierre Curie: published by the French Physical Society}}}\ (\bibinfo  {publisher} {Gauthier-Villars},\ \bibinfo {year} {1908})\BibitemShut {NoStop}%
\bibitem [{\citenamefont {Arfken}, \citenamefont {Weber},\ and\ \citenamefont {Harris}(2012)}]{arfken12}%
  \BibitemOpen
  \bibfield  {author} {\bibinfo {author} {\bibfnamefont {G.~B.}\ \bibnamefont {Arfken}}, \bibinfo {author} {\bibfnamefont {H.-J.}\ \bibnamefont {Weber}}, \ and\ \bibinfo {author} {\bibfnamefont {F.~E.}\ \bibnamefont {Harris}},\ }\href@noop {} {\emph {\bibinfo {title} {Mathematical Methods for Physicists}}}\ (\bibinfo  {publisher} {Academic},\ \bibinfo {address} {Oxford},\ \bibinfo {year} {2012})\BibitemShut {NoStop}%
\bibitem [{\citenamefont {Hooyman}, \citenamefont {Mazur},\ and\ \citenamefont {{de Groot}}(1954)}]{hooyman_1954}%
  \BibitemOpen
  \bibfield  {author} {\bibinfo {author} {\bibfnamefont {G.}~\bibnamefont {Hooyman}}, \bibinfo {author} {\bibfnamefont {P.}~\bibnamefont {Mazur}}, \ and\ \bibinfo {author} {\bibfnamefont {S.}~\bibnamefont {{de Groot}}},\ }\href {\doibase https://doi.org/10.1016/S0031-8914(54)91998-9} {\bibfield  {journal} {\bibinfo  {journal} {Physica}\ }\textbf {\bibinfo {volume} {21}},\ \bibinfo {pages} {355} (\bibinfo {year} {1954})}\BibitemShut {NoStop}%
\bibitem [{\citenamefont {Scheiner}\ and\ \citenamefont {Baalrud}(2020)}]{Scheiner_2020}%
  \BibitemOpen
  \bibfield  {author} {\bibinfo {author} {\bibfnamefont {B.}~\bibnamefont {Scheiner}}\ and\ \bibinfo {author} {\bibfnamefont {S.~D.}\ \bibnamefont {Baalrud}},\ }\href {\doibase 10.1103/PhysRevE.102.063202} {\bibfield  {journal} {\bibinfo  {journal} {Phys. Rev. E}\ }\textbf {\bibinfo {volume} {102}},\ \bibinfo {pages} {063202} (\bibinfo {year} {2020})}\BibitemShut {NoStop}%
\bibitem [{\citenamefont {Istomin}\ and\ \citenamefont {Kustova}(2017)}]{Istomin_2017}%
  \BibitemOpen
  \bibfield  {author} {\bibinfo {author} {\bibfnamefont {V.~A.}\ \bibnamefont {Istomin}}\ and\ \bibinfo {author} {\bibfnamefont {E.~V.}\ \bibnamefont {Kustova}},\ }\href {\doibase 10.1063/1.4975315} {\bibfield  {journal} {\bibinfo  {journal} {Physics of Plasmas}\ }\textbf {\bibinfo {volume} {24}},\ \bibinfo {pages} {022109} (\bibinfo {year} {2017})}\BibitemShut {NoStop}%
\bibitem [{\citenamefont {Ferziger}\ and\ \citenamefont {Kaper}(1972)}]{ferziger_kaper}%
  \BibitemOpen
  \bibfield  {author} {\bibinfo {author} {\bibfnamefont {J.}~\bibnamefont {Ferziger}}\ and\ \bibinfo {author} {\bibfnamefont {H.}~\bibnamefont {Kaper}},\ }\href {https://books.google.com/books?id=mxS2AAAAIAAJ} {\emph {\bibinfo {title} {Mathematical Theory of Transport Processes in Gases}}}\ (\bibinfo  {publisher} {North-Holland Publishing Company},\ \bibinfo {year} {1972})\BibitemShut {NoStop}%
\bibitem [{\citenamefont {Hansen}\ and\ \citenamefont {McDonald}(2013)}]{hansen_2013}%
  \BibitemOpen
  \bibfield  {author} {\bibinfo {author} {\bibfnamefont {J.-P.}\ \bibnamefont {Hansen}}\ and\ \bibinfo {author} {\bibfnamefont {I.~R.}\ \bibnamefont {McDonald}},\ }\href@noop {} {\emph {\bibinfo {title} {Theory of simple liquids: with applications to soft matter}}}\ (\bibinfo  {publisher} {Academic press},\ \bibinfo {year} {2013})\BibitemShut {NoStop}%
\bibitem [{\citenamefont {Reif}(1965)}]{Reif_1965}%
  \BibitemOpen
  \bibfield  {author} {\bibinfo {author} {\bibfnamefont {F.}~\bibnamefont {Reif}},\ }\href@noop {} {\emph {\bibinfo {title} {Fundamentals of Statistical and Thermal Physics}}}\ (\bibinfo  {publisher} {McGraw Hill},\ \bibinfo {address} {Tokyo},\ \bibinfo {year} {1965})\BibitemShut {NoStop}%
\bibitem [{\citenamefont {Pathria}\ and\ \citenamefont {Beale}(2011)}]{pathria2011}%
  \BibitemOpen
  \bibfield  {author} {\bibinfo {author} {\bibfnamefont {R.~K.}\ \bibnamefont {Pathria}}\ and\ \bibinfo {author} {\bibfnamefont {P.~D.}\ \bibnamefont {Beale}},\ }\href@noop {} {\emph {\bibinfo {title} {Statistical Mechanics}}},\ \bibinfo {edition} {3rd}\ ed.\ (\bibinfo  {publisher} {Elsevier/Academic Press},\ \bibinfo {address} {Amsterdam ; Boston},\ \bibinfo {year} {2011})\BibitemShut {NoStop}%
\bibitem [{\citenamefont {Ichimaru}(1991)}]{ichimaru_book}%
  \BibitemOpen
  \bibfield  {author} {\bibinfo {author} {\bibnamefont {Ichimaru}},\ }\href@noop {} {\emph {\bibinfo {title} {Statistical Plasma Physics Volume I}}}\ (\bibinfo  {publisher} {CRC Press,},\ \bibinfo {year} {1991})\BibitemShut {NoStop}%
\bibitem [{\citenamefont {J~Evans}\ and\ \citenamefont {P~Morriss}(2007)}]{Evans_Morriss}%
  \BibitemOpen
  \bibfield  {author} {\bibinfo {author} {\bibfnamefont {D.}~\bibnamefont {J~Evans}}\ and\ \bibinfo {author} {\bibfnamefont {G.}~\bibnamefont {P~Morriss}},\ }\href@noop {} {\emph {\bibinfo {title} {Statistical mechanics of nonequilbrium liquids}}}\ (\bibinfo  {publisher} {ANU Press},\ \bibinfo {year} {2007})\BibitemShut {NoStop}%
\bibitem [{\citenamefont {Vieillefosse}\ and\ \citenamefont {Hansen}(1975)}]{HansenPRA1975}%
  \BibitemOpen
  \bibfield  {author} {\bibinfo {author} {\bibfnamefont {P.}~\bibnamefont {Vieillefosse}}\ and\ \bibinfo {author} {\bibfnamefont {J.~P.}\ \bibnamefont {Hansen}},\ }\href {\doibase 10.1103/PhysRevA.12.1106} {\bibfield  {journal} {\bibinfo  {journal} {Phys. Rev. A}\ }\textbf {\bibinfo {volume} {12}},\ \bibinfo {pages} {1106} (\bibinfo {year} {1975})}\BibitemShut {NoStop}%
\bibitem [{\citenamefont {Ott}, \citenamefont {Bonitz},\ and\ \citenamefont {Donk\'o}(2015)}]{OttPRE2015}%
  \BibitemOpen
  \bibfield  {author} {\bibinfo {author} {\bibfnamefont {T.}~\bibnamefont {Ott}}, \bibinfo {author} {\bibfnamefont {M.}~\bibnamefont {Bonitz}}, \ and\ \bibinfo {author} {\bibfnamefont {Z.}~\bibnamefont {Donk\'o}},\ }\href {\doibase 10.1103/PhysRevE.92.063105} {\bibfield  {journal} {\bibinfo  {journal} {Phys. Rev. E}\ }\textbf {\bibinfo {volume} {92}},\ \bibinfo {pages} {063105} (\bibinfo {year} {2015})}\BibitemShut {NoStop}%
\bibitem [{\citenamefont {Donk\'o}\ \emph {et~al.}(1998)\citenamefont {Donk\'o}, \citenamefont {Ny\'{\i}ri}, \citenamefont {Szalai},\ and\ \citenamefont {Holl\'o}}]{DonkoPRL1998}%
  \BibitemOpen
  \bibfield  {author} {\bibinfo {author} {\bibfnamefont {Z.}~\bibnamefont {Donk\'o}}, \bibinfo {author} {\bibfnamefont {B.}~\bibnamefont {Ny\'{\i}ri}}, \bibinfo {author} {\bibfnamefont {L.}~\bibnamefont {Szalai}}, \ and\ \bibinfo {author} {\bibfnamefont {S.}~\bibnamefont {Holl\'o}},\ }\href {\doibase 10.1103/PhysRevLett.81.1622} {\bibfield  {journal} {\bibinfo  {journal} {Phys. Rev. Lett.}\ }\textbf {\bibinfo {volume} {81}},\ \bibinfo {pages} {1622} (\bibinfo {year} {1998})}\BibitemShut {NoStop}%
\bibitem [{\citenamefont {Daligault}, \citenamefont {Rasmussen},\ and\ \citenamefont {Baalrud}(2014)}]{DaligaultPRE2014}%
  \BibitemOpen
  \bibfield  {author} {\bibinfo {author} {\bibfnamefont {J.}~\bibnamefont {Daligault}}, \bibinfo {author} {\bibfnamefont {K.~O.}\ \bibnamefont {Rasmussen}}, \ and\ \bibinfo {author} {\bibfnamefont {S.~D.}\ \bibnamefont {Baalrud}},\ }\href {\doibase 10.1103/PhysRevE.90.033105} {\bibfield  {journal} {\bibinfo  {journal} {Phys. Rev. E}\ }\textbf {\bibinfo {volume} {90}},\ \bibinfo {pages} {033105} (\bibinfo {year} {2014})}\BibitemShut {NoStop}%
\bibitem [{\citenamefont {Zwanzig}(1964)}]{Zwanzig_1964}%
  \BibitemOpen
  \bibfield  {author} {\bibinfo {author} {\bibfnamefont {R.}~\bibnamefont {Zwanzig}},\ }\href {\doibase 10.1063/1.1725558} {\bibfield  {journal} {\bibinfo  {journal} {The Journal of Chemical Physics}\ }\textbf {\bibinfo {volume} {40}},\ \bibinfo {pages} {2527} (\bibinfo {year} {1964})}\BibitemShut {NoStop}%
\bibitem [{\citenamefont {Frenkel}\ and\ \citenamefont {Smit}(2002)}]{Frenkel_book}%
  \BibitemOpen
  \bibfield  {author} {\bibinfo {author} {\bibfnamefont {D.}~\bibnamefont {Frenkel}}\ and\ \bibinfo {author} {\bibfnamefont {B.}~\bibnamefont {Smit}},\ }\href@noop {} {\emph {\bibinfo {title} {Understanding Molecular Simulation: From Algorithms to Applications}}},\ \bibinfo {edition} {2nd}\ ed.,\ \bibinfo {series} {Computational Science Series}, Vol.~\bibinfo {volume} {1}\ (\bibinfo  {publisher} {Academic Press},\ \bibinfo {address} {San Diego},\ \bibinfo {year} {2002})\BibitemShut {NoStop}%
\bibitem [{\citenamefont {Glenzer}\ and\ \citenamefont {Redmer}(2009)}]{GlenzerRMP2009}%
  \BibitemOpen
  \bibfield  {author} {\bibinfo {author} {\bibfnamefont {S.~H.}\ \bibnamefont {Glenzer}}\ and\ \bibinfo {author} {\bibfnamefont {R.}~\bibnamefont {Redmer}},\ }\href {\doibase 10.1103/RevModPhys.81.1625} {\bibfield  {journal} {\bibinfo  {journal} {Rev. Mod. Phys.}\ }\textbf {\bibinfo {volume} {81}},\ \bibinfo {pages} {1625} (\bibinfo {year} {2009})}\BibitemShut {NoStop}%
\bibitem [{\citenamefont {Brush}, \citenamefont {Sahlin},\ and\ \citenamefont {Teller}(1966)}]{BrushJCP1966}%
  \BibitemOpen
  \bibfield  {author} {\bibinfo {author} {\bibfnamefont {S.~G.}\ \bibnamefont {Brush}}, \bibinfo {author} {\bibfnamefont {H.~L.}\ \bibnamefont {Sahlin}}, \ and\ \bibinfo {author} {\bibfnamefont {E.}~\bibnamefont {Teller}},\ }\href {\doibase 10.1063/1.1727895} {\bibfield  {journal} {\bibinfo  {journal} {The Journal of Chemical Physics}\ }\textbf {\bibinfo {volume} {45}},\ \bibinfo {pages} {2102} (\bibinfo {year} {1966})},\ \Eprint {http://arxiv.org/abs/https://pubs.aip.org/aip/jcp/article-pdf/45/6/2102/18845826/2102\_1\_online.pdf} {https://pubs.aip.org/aip/jcp/article-pdf/45/6/2102/18845826/2102\_1\_online.pdf} \BibitemShut {NoStop}%
\bibitem [{\citenamefont {Baus}\ and\ \citenamefont {Hansen}(1980)}]{Baus_1980}%
  \BibitemOpen
  \bibfield  {author} {\bibinfo {author} {\bibfnamefont {M.}~\bibnamefont {Baus}}\ and\ \bibinfo {author} {\bibfnamefont {J.-P.}\ \bibnamefont {Hansen}},\ }\href {\doibase https://doi.org/10.1016/0370-1573(80)90022-8} {\bibfield  {journal} {\bibinfo  {journal} {Physics Reports}\ }\textbf {\bibinfo {volume} {59}},\ \bibinfo {pages} {1} (\bibinfo {year} {1980})}\BibitemShut {NoStop}%
\bibitem [{\citenamefont {Tsednee}\ and\ \citenamefont {Luchko}(2019)}]{Tsednee_2019}%
  \BibitemOpen
  \bibfield  {author} {\bibinfo {author} {\bibfnamefont {T.}~\bibnamefont {Tsednee}}\ and\ \bibinfo {author} {\bibfnamefont {T.}~\bibnamefont {Luchko}},\ }\href {\doibase 10.1103/PhysRevE.99.032130} {\bibfield  {journal} {\bibinfo  {journal} {Phys. Rev. E}\ }\textbf {\bibinfo {volume} {99}},\ \bibinfo {pages} {032130} (\bibinfo {year} {2019})}\BibitemShut {NoStop}%
\bibitem [{\citenamefont {Filinov}\ \emph {et~al.}(2004)\citenamefont {Filinov}, \citenamefont {Golubnychiy}, \citenamefont {Bonitz}, \citenamefont {Ebeling},\ and\ \citenamefont {Dufty}}]{filinov2004temperature}%
  \BibitemOpen
  \bibfield  {author} {\bibinfo {author} {\bibfnamefont {A.~V.}\ \bibnamefont {Filinov}}, \bibinfo {author} {\bibfnamefont {V.~O.}\ \bibnamefont {Golubnychiy}}, \bibinfo {author} {\bibfnamefont {M.}~\bibnamefont {Bonitz}}, \bibinfo {author} {\bibfnamefont {W.}~\bibnamefont {Ebeling}}, \ and\ \bibinfo {author} {\bibfnamefont {J.~W.}\ \bibnamefont {Dufty}},\ }\href {\doibase 10.1103/PhysRevE.70.046411} {\bibfield  {journal} {\bibinfo  {journal} {Phys. Rev. E}\ }\textbf {\bibinfo {volume} {70}},\ \bibinfo {pages} {046411} (\bibinfo {year} {2004})}\BibitemShut {NoStop}%
\bibitem [{\citenamefont {Dornheim}\ \emph {et~al.}(2025)\citenamefont {Dornheim}, \citenamefont {Bonitz}, \citenamefont {Moldabekov}, \citenamefont {Schwalbe}, \citenamefont {Tolias},\ and\ \citenamefont {Vorberger}}]{DornheimPRB2025}%
  \BibitemOpen
  \bibfield  {author} {\bibinfo {author} {\bibfnamefont {T.}~\bibnamefont {Dornheim}}, \bibinfo {author} {\bibfnamefont {M.}~\bibnamefont {Bonitz}}, \bibinfo {author} {\bibfnamefont {Z.~A.}\ \bibnamefont {Moldabekov}}, \bibinfo {author} {\bibfnamefont {S.}~\bibnamefont {Schwalbe}}, \bibinfo {author} {\bibfnamefont {P.}~\bibnamefont {Tolias}}, \ and\ \bibinfo {author} {\bibfnamefont {J.}~\bibnamefont {Vorberger}},\ }\href {\doibase 10.1103/PhysRevB.111.115149} {\bibfield  {journal} {\bibinfo  {journal} {Phys. Rev. B}\ }\textbf {\bibinfo {volume} {111}},\ \bibinfo {pages} {115149} (\bibinfo {year} {2025})}\BibitemShut {NoStop}%
\bibitem [{\citenamefont {Bonitz}\ \emph {et~al.}(2024)\citenamefont {Bonitz}, \citenamefont {Vorberger}, \citenamefont {Bethkenhagen}, \citenamefont {Böhme}, \citenamefont {Ceperley}, \citenamefont {Filinov}, \citenamefont {Gawne}, \citenamefont {Graziani}, \citenamefont {Gregori}, \citenamefont {Hamann}, \citenamefont {Hansen}, \citenamefont {Holzmann}, \citenamefont {Hu}, \citenamefont {Kählert}, \citenamefont {Karasiev}, \citenamefont {Kleinschmidt}, \citenamefont {Kordts}, \citenamefont {Makait}, \citenamefont {Militzer}, \citenamefont {Moldabekov}, \citenamefont {Pierleoni}, \citenamefont {Preising}, \citenamefont {Ramakrishna}, \citenamefont {Redmer}, \citenamefont {Schwalbe}, \citenamefont {Svensson},\ and\ \citenamefont {Dornheim}}]{BonitzPOP2024}%
  \BibitemOpen
  \bibfield  {author} {\bibinfo {author} {\bibfnamefont {M.}~\bibnamefont {Bonitz}}, \bibinfo {author} {\bibfnamefont {J.}~\bibnamefont {Vorberger}}, \bibinfo {author} {\bibfnamefont {M.}~\bibnamefont {Bethkenhagen}}, \bibinfo {author} {\bibfnamefont {M.~P.}\ \bibnamefont {Böhme}}, \bibinfo {author} {\bibfnamefont {D.~M.}\ \bibnamefont {Ceperley}}, \bibinfo {author} {\bibfnamefont {A.}~\bibnamefont {Filinov}}, \bibinfo {author} {\bibfnamefont {T.}~\bibnamefont {Gawne}}, \bibinfo {author} {\bibfnamefont {F.}~\bibnamefont {Graziani}}, \bibinfo {author} {\bibfnamefont {G.}~\bibnamefont {Gregori}}, \bibinfo {author} {\bibfnamefont {P.}~\bibnamefont {Hamann}}, \bibinfo {author} {\bibfnamefont {S.~B.}\ \bibnamefont {Hansen}}, \bibinfo {author} {\bibfnamefont {M.}~\bibnamefont {Holzmann}}, \bibinfo {author} {\bibfnamefont {S.~X.}\ \bibnamefont {Hu}}, \bibinfo {author} {\bibfnamefont {H.}~\bibnamefont {Kählert}}, \bibinfo {author} {\bibfnamefont {V.~V.}\ \bibnamefont {Karasiev}}, \bibinfo {author} {\bibfnamefont
  {U.}~\bibnamefont {Kleinschmidt}}, \bibinfo {author} {\bibfnamefont {L.}~\bibnamefont {Kordts}}, \bibinfo {author} {\bibfnamefont {C.}~\bibnamefont {Makait}}, \bibinfo {author} {\bibfnamefont {B.}~\bibnamefont {Militzer}}, \bibinfo {author} {\bibfnamefont {Z.~A.}\ \bibnamefont {Moldabekov}}, \bibinfo {author} {\bibfnamefont {C.}~\bibnamefont {Pierleoni}}, \bibinfo {author} {\bibfnamefont {M.}~\bibnamefont {Preising}}, \bibinfo {author} {\bibfnamefont {K.}~\bibnamefont {Ramakrishna}}, \bibinfo {author} {\bibfnamefont {R.}~\bibnamefont {Redmer}}, \bibinfo {author} {\bibfnamefont {S.}~\bibnamefont {Schwalbe}}, \bibinfo {author} {\bibfnamefont {P.}~\bibnamefont {Svensson}}, \ and\ \bibinfo {author} {\bibfnamefont {T.}~\bibnamefont {Dornheim}},\ }\href {\doibase 10.1063/5.0219405} {\bibfield  {journal} {\bibinfo  {journal} {Physics of Plasmas}\ }\textbf {\bibinfo {volume} {31}},\ \bibinfo {pages} {110501} (\bibinfo {year} {2024})},\ \Eprint
  {http://arxiv.org/abs/https://pubs.aip.org/aip/pop/article-pdf/doi/10.1063/5.0219405/20250831/110501\_1\_5.0219405.pdf} {https://pubs.aip.org/aip/pop/article-pdf/doi/10.1063/5.0219405/20250831/110501\_1\_5.0219405.pdf} \BibitemShut {NoStop}%
\bibitem [{\citenamefont {van Leeuwen}(1921)}]{van_leeuwen}%
  \BibitemOpen
  \bibfield  {author} {\bibinfo {author} {\bibfnamefont {H.-J.}\ \bibnamefont {van Leeuwen}},\ }\href {\doibase 10.1051/jphysrad:01921002012036100} {\bibfield  {journal} {\bibinfo  {journal} {{Journal de Physique et le Radium}}\ }\textbf {\bibinfo {volume} {2}},\ \bibinfo {pages} {361} (\bibinfo {year} {1921})}\BibitemShut {NoStop}%
\bibitem [{\citenamefont {Balescu}(1988)}]{balescu_text}%
  \BibitemOpen
  \bibfield  {author} {\bibinfo {author} {\bibfnamefont {R.}~\bibnamefont {Balescu}},\ }in\ \href {\doibase https://doi.org/10.1016/B978-0-444-87091-9.50009-9} {\emph {\bibinfo {booktitle} {Classical Transport}}},\ \bibinfo {series and number} {Transport Processes in Plasmas},\ \bibinfo {editor} {edited by\ \bibinfo {editor} {\bibfnamefont {R.}~\bibnamefont {Balescu}}}\ (\bibinfo  {publisher} {North-Holland},\ \bibinfo {address} {Amsterdam},\ \bibinfo {year} {1988})\ pp.\ \bibinfo {pages} {211--276}\BibitemShut {NoStop}%
\bibitem [{\citenamefont {Gurnett}\ and\ \citenamefont {Bhattacharjee}(2005)}]{gurnett_2005}%
  \BibitemOpen
  \bibfield  {author} {\bibinfo {author} {\bibfnamefont {D.~A.}\ \bibnamefont {Gurnett}}\ and\ \bibinfo {author} {\bibfnamefont {A.}~\bibnamefont {Bhattacharjee}},\ }\href@noop {} {\emph {\bibinfo {title} {Introduction to plasma physics: with space and laboratory applications}}}\ (\bibinfo  {publisher} {Cambridge university press},\ \bibinfo {year} {2005})\BibitemShut {NoStop}%
\bibitem [{\citenamefont {Boyd}\ and\ \citenamefont {Sanderson}(2003)}]{Boyd_Sanderson_2003}%
  \BibitemOpen
  \bibfield  {author} {\bibinfo {author} {\bibfnamefont {T.~J.~M.}\ \bibnamefont {Boyd}}\ and\ \bibinfo {author} {\bibfnamefont {J.~J.}\ \bibnamefont {Sanderson}},\ }\href@noop {} {\emph {\bibinfo {title} {The Physics of Plasmas}}}\ (\bibinfo  {publisher} {Cambridge University Press},\ \bibinfo {year} {2003})\BibitemShut {NoStop}%
\bibitem [{\citenamefont {Freidberg}(2014)}]{Freidberg_book}%
  \BibitemOpen
  \bibfield  {author} {\bibinfo {author} {\bibfnamefont {J.~P.}\ \bibnamefont {Freidberg}},\ }\href@noop {} {\emph {\bibinfo {title} {Ideal MHD}}}\ (\bibinfo  {publisher} {Cambridge University Press},\ \bibinfo {year} {2014})\BibitemShut {NoStop}%
\end{thebibliography}%

\end{document}